\documentclass[twocolumn]{aastex63}
\usepackage{color}

\usepackage{multirow}
\usepackage{amsmath}

\shortauthors{Fukui, Sano, Yamane et al.}

\begin{document}
\title{Pursuing the origin of the gamma rays in RX~J1713.7$-$3946 \\quantifying the hadronic and leptonic components}

\author[0000-0002-8966-9856]{Yasuo Fukui}
\affiliation{Department of Physics, Nagoya University, Furo-cho, Chikusa-ku, Nagoya 464-8601, Japan; fukui@a.phys.nagoya-u.ac.jp}

\author[0000-0003-2062-5692]{Hidetoshi Sano}
\affiliation{Department of Physics, Nagoya University, Furo-cho, Chikusa-ku, Nagoya 464-8601, Japan; fukui@a.phys.nagoya-u.ac.jp}
\affiliation{Division of Science, National Astronomical Observatory of Japan, Mitaka, Tokyo 181-8588, Japan; hidetoshi.sano@nao.ac.jp}

\author[0000-0001-8296-7482]{Yumiko Yamane}
\affiliation{Department of Physics, Nagoya University, Furo-cho, Chikusa-ku, Nagoya 464-8601, Japan; fukui@a.phys.nagoya-u.ac.jp}

\author[0000-0003-0324-1689]{Takahiro Hayakawa}
\affiliation{Department of Physics, Nagoya University, Furo-cho, Chikusa-ku, Nagoya 464-8601, Japan; fukui@a.phys.nagoya-u.ac.jp}

\author[0000-0002-7935-8771]{Tsuyoshi Inoue}
\affiliation{Department of Physics, Nagoya University, Furo-cho, Chikusa-ku, Nagoya 464-8601, Japan; fukui@a.phys.nagoya-u.ac.jp}

\author[0000-0002-1411-5410]{Kengo Tachihara}
\affiliation{Department of Physics, Nagoya University, Furo-cho, Chikusa-ku, Nagoya 464-8601, Japan; fukui@a.phys.nagoya-u.ac.jp}

\author[0000-0002-9516-1581]{Gavin Rowell}
\affiliation{School of Physical Sciences, The University of Adelaide, North Terrace, Adelaide, SA 5005, Australia}

\author[0000-0001-9687-8237]{Sabrina Einecke}
\affiliation{School of Physical Sciences, The University of Adelaide, North Terrace, Adelaide, SA 5005, Australia}

\begin{abstract}
We analyzed the TeV gamma-ray image of a supernova remnant RX~J1713.7$-$3946 (RX~J1713) through a comparison with the interstellar medium (ISM) and the non-thermal X-rays. The gamma-ray datasets at two energy bands of $>$2~TeV and $>$250--300~GeV were obtained with H.E.S.S. \citep{2018A+A...612A...6H,2007AA...464..235A} and utilized in the analysis. We employed a new methodology which assumes that the gamma-ray counts are expressed by a linear combination of two terms; one is proportional to the ISM column density and the other proportional to the X-ray count. We then assume these represent the hadronic and leptonic  components, respectively. By fitting the expression to the data pixels, we find that the gamma-ray counts are well represented by a flat plane in a 3D space of the gamma-ray counts, the ISM column density and the X-ray counts. The results using the latest H.E.S.S. data at 4\farcm8 resolution show that the hadronic and leptonic components occupy $(67\pm8)$\% and $(33\pm8)$\% of the total gamma rays, respectively, where the two components have been quantified for the first time. The hadronic component is greater than the leptonic component, which reflects the massive ISM of $\sim$10$^4$~$M_{\odot}$ associated with the SNR, lending support for the acceleration of the cosmic-ray protons. There is a marginal hint that the gamma rays are suppressed at high gamma-ray counts which may be ascribed to the second order effects including the shock-cloud interaction and the penetration-depth effect.
\end{abstract}
\vspace*{-0.3cm}

\section{Introduction} \label{sec:intro}
RX~J1713.7$-$3946 (hereafter referred to as RX~J1713) is the brightest TeV gamma-ray and non-thermal X-ray supernova remnant (SNR) as revealed by {the H.E.S.S. telescope array} \citep[][]{2004Natur.432...75A, 2006A&A...449..223A} and {\textit{ROSAT}} X-ray observations \citep{1996rftu.proc..267P}, and has been the primary target where the gamma-ray origin can be established. The gamma-rays are produced either by the leptonic origin or the hadronic origin in a SNR. In the leptonic process cosmic-ray (CR) electrons energize the low-energy photons into the gamma-rays, and in the hadronic process CR protons react with the interstellar protons to produce neutral pions which decay into two gamma-rays. If the hadronic process is established, it provides evidence for CR acceleration in a SNR. Considerable effort {has} been made in modeling the broad band radiation spectrum in the SNR \citep[e.g.,][see also the references therein]{2006A&A...449..223A, 2018A+A...612A...6H}, and it is shown that either of the two mechanisms can reproduce the observed spectrum. It still remains unsettled what is the {dominant} gamma-ray production mechanism in RX~J1713. It is {also} shown that {leptonic} Bremsstrahlung is not effective \citep[e.g.,][hereafter ZA10]{2010ApJ...708..965Z} \defcitealias{2010ApJ...708..965Z}{ZA10} and is not considered in the present work.

\citetalias{2010ApJ...708..965Z} (see also references therein) developed detailed theoretical modelling of the high and very high energy {(HE and VHE)} radiation of the SNR and presented models of the radiation spectrum in the diffusive shock acceleration (DSA) scheme of CRs \citep[e.g.,][]{1978MNRAS.182..147B, 1978ApJ...221L..29B}. These authors discussed the leptonic and hadronic origins of the gamma-ray production as well as {a} composite origin by referring to the CO dense cores overtaken by the SNR shock (\citealt[F03 hereafter]{2003PASJ...55L..61F}; \citealt{2008AIPC.1085..104F}),\defcitealias{2003PASJ...55L..61F}{F03} where the CO emission is used as proxy {for} H$_2$. Other theoretical studies \citep[e.g.,][]{2010ApJ...712..287E} argued for the leptonic process, because the gas density, if {assumed to be uniform}, required to produce gamma rays via the hadronic process will produce significant thermal X-rays, which contradicts the observed purely non-thermal X-rays (\citealt{1997PASJ...49L...7K}; \citealt{2008ApJ...685..988T}; for thermal X-rays observed in a local spot, {\citep[see also][]{2015ApJ...814...29K}}. The non-thermal X-rays show spatial distribution similar to the gamma rays, which may suggest the leptonic component \citep[e.g.,][]{2006A&A...449..223A, 2018A+A...612A...6H}.

\citetalias{2003PASJ...55L..61F} identified the CO gas at $\sim -10$\,km\,s$^{-1}$ interacting with RX~J1713 by observations with the NANTEN telescope, and determined a kinematical distance of the SNR to be 1\,kpc {(see the appendix \ref{sec:app:distance} for more detail)}. {This value is consistent with the latest estimation of $1.12 \pm 0.01$ kpc \citep[][]{2021NatAs.tmp...79L}}. \citetalias{2003PASJ...55L..61F} further found that a CO cloud named peak D, having a size of $\sim 5\arcmin\times10\arcmin$ \citepalias[see Figure 2 of][]{2003PASJ...55L..61F}, shows a good coincidence with the TeV gamma ray peak, and suggested that the peak D cloud plays a role of target protons in the hadronic gamma-ray production. Subsequently, \citet{2005ApJ...631..947M} and \citet{2013ApJ...778...59S} extended a list of CO peaks in RX~J1713 based on the NANTEN observations. \citet{2006A&A...449..223A} compared the NANTEN CO distribution with the H.E.S.S. TeV gamma-ray image and examined both leptonic and hadronic scenarios as the origin of the gamma rays. These authors found that TeV gamma rays show a similar trend to the CO distribution while their correspondence is not good {especially} in the south- eastern rim of the TeV gamma-ray shell. \citet[][F12 hereafter]{2012ApJ...746...82F}\defcitealias{2012ApJ...746...82F}{F12} {thought} that not only H$_2$ but also \ion{H}{1} is important as target protons {for} the hadronic process. These authors showed that the total interstellar medium (ISM) {in RX~J1713} consists of both atomic and molecular protons, and demonstrated that the ISM shows a good spatial correspondence with the TeV gamma rays observed with H.E.S.S. \citetalias{2012ApJ...746...82F} interpreted {this as support for} the hadronic origin, while the effective spatial resolution employed $\sim$4\,pc, was not sufficiently high to resolve the degeneracy between the gamma-ray and X-ray distributions, leaving room for leptonic gamma rays.

The Fermi collaboration \citep{2011ApJ...734...28A} {revealed} a hard gamma-ray spectrum in RX~J1713 and argued for the leptonic gamma-rays by assuming uniform ISM distribution. \citet[][IYIF12 {hereafter}]{2012ApJ...744...71I}\defcitealias{2012ApJ...744...71I}{IYIF12} carried out {magneto-hydrodynamics (MHD)} simulations of the shock-cloud interaction in the realistic clumpy ISM distribution and showed that the interaction makes the magnetic field highly turbulent with field strength up to $\sim$1\,mG around dense cores. Furthermore, \citetalias{2012ApJ...744...71I} derived an expression of the penetration depth of CRs into dense cores and argued that the energy-dependent penetration depth hardens the hadronic gamma-ray spectrum toward the dense cores into one similar to the spectrum observed with Fermi. Prior to \citetalias{2012ApJ...744...71I}, the effect of the penetration depth in dense cores was pointed out in terms of the CR diffusion coefficient by \citet{2007Ap&SS.309..365G}, while hydrodynamical simulations of the shock interaction were not undertaken. \citetalias{2010ApJ...708..965Z} also suggested hardening of the gamma-ray spectrum in the clumpy ISM qualitatively, and \citet{2014MNRAS.445L..70G} demonstrated the hard gamma-ray spectrum in clumpy ISM by considering the CR penetration limited by the diffusion coefficient. Most recently, the new H.E.S.S. gamma-ray data were released, where the sensitivity was improved by a factor of two at three-time better spatial resolution {\citep{2018A+A...612A...6H}} than the previous H.E.S.S. data \citep{2006A&A...449..223A}. The new spatial resolution 1.4\,pc at $>2$\,TeV is still not {good} enough to fully resolve the gamma-ray/X-ray degeneracy, and the debate on the two origins is {ongoing}.

\begin{figure*}
\begin{center}
\includegraphics[width=160mm,clip]{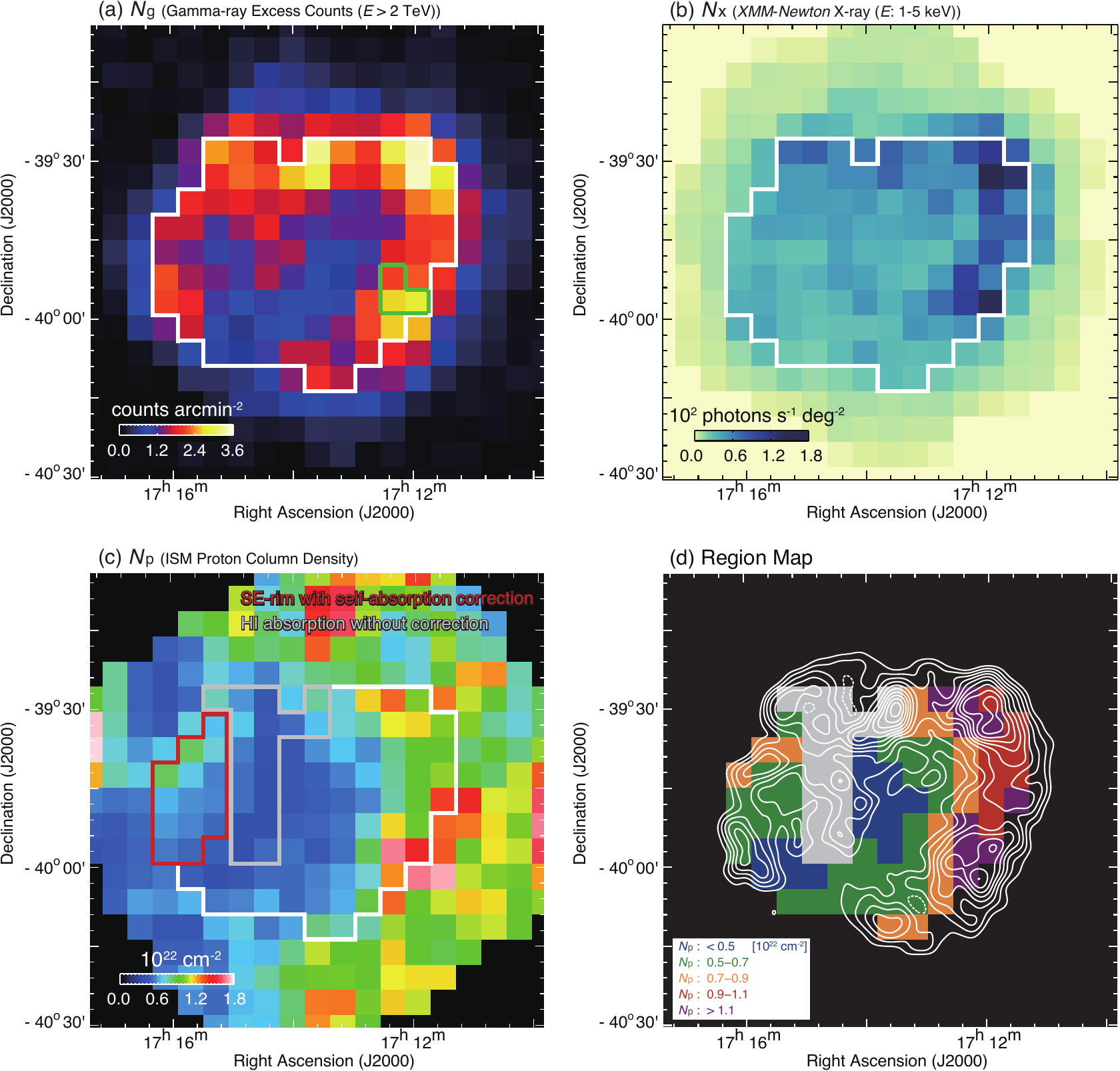}
\caption{{Spatial distributions of (a) $N_{\rm g}$ (H.E.S.S. TeV gamma-rays \citep[$E > 2$ TeV,][]{2018A+A...612A...6H}, (b) $N_{\rm x}$ ({\it{XMM-Newton}} X-rays {($E$: 1.0--5.0\,keV)}), and (c) $N_{\rm p}$  \citepalias[ISM proton column density,][]{2012ApJ...746...82F}. {Three pixels within the green lines are discussed in Section \ref{2ndF}.} White {polygon} indicates region to be used for the present analysis. Red {polygon} indicates region where we make correction for the \ion{H}{1} absorption (see Section \ref{obs:ISM}). The pixels within the gray polygon represents region where there is no reliable way to assume background \ion{H}{1} without self-absorption and are excluded in the present analysis {(see text)}. (d) Gray {pixels} are the same as region within the gray {polygon.} Blue, green, orange, red, and purple pixels represent the regions where $N_{\rm p}$ less than 0.5, 0.5--0.7, 0.7--0.9, 0.9--1.1, more than $1.1\times 10^{22}$\,cm$^{-2}$, respectively. Superposed {white} contours indicate TeV gamma-rays excess counts. The lowest contour level and contour intervals are {12} and {3} excess counts, respectively.}}
\label{fig1}
\end{center}
\end{figure*}

It is generally expected that the observed gamma rays are combination of the leptonic and hadronic gamma rays, where the former is proportional to the CR electrons and the low-energy photons and the latter to the CR protons and the ISM protons \citep[e.g.,][]{1994A&A...285..645A, 1994A&A...287..959D, 2010ApJ...708..965Z}. The CR electrons are to a first approximation proportional to the non-thermal X-rays {for the inverse Compton scattering} in the uniform background photon field (e.g., the Cosmic Microwave Background), while {the} shock-cloud interactions may vary the relationship through the amplification of the magnetic field, especially around the dense cores \citepalias{2012ApJ...744...71I}. Such amplification became clear through enhanced X-rays as revealed observationally by \citet{2010ApJ...724...59S, 2013ApJ...778...59S}. Furthermore, the penetration depth of CRs may become limited toward dense cores, leading to {suppress the} hadronic gamma-rays \citep{2007Ap&SS.309..365G}. The effect of the penetration depth has not been directly shown observationally except for the GeV gamma-ray spectrum by Fermi LAT \citepalias[e.g.,][]{2012ApJ...744...71I}, while spatial anti-correlation between dense cores and gamma rays may  become apparent if angular resolution becomes sufficiently high. It is in particular an important challenge to discern the two gamma-ray origins, and the new H.E.S.S. data may offer an opportunity for such an attempt.

\begin{deluxetable*}{lrrccrrr}
\tablewidth{\linewidth}
\tablecaption{Summary of $XMM$-$Newton$ archive data in RX~J1713.7$-$3946}
\tablehead{
&&&&&\multicolumn{3}{c}{Exposure}\\
\cline{6-8}\\
\colhead{Observation ID} & \colhead{$\alpha_{\mathrm{J2000}}$} & \colhead{$\delta_{\mathrm{J2000}}$}  & \colhead{Start Date} & \colhead{End Date} & \colhead{MOS1} & \colhead{MOS2} & \colhead{pn} \\
& \colhead{(degree)} & \colhead{(degree)} & \colhead{(yyyy-mm-dd hh:mm:ss)} & \colhead{(yyyy-mm-dd hh:mm:ss)} & \colhead{(ks)} & \colhead{(ks)} & \colhead{(ks)} \\
}
\startdata
0093670101 & 258.55 & $-39.43$ & 2001-09-05 03:42:27 & 2001-09-05 06:29:06 & 1.7 & 1.7 & 9.9 \\
0093670201 & 258.00 & $-39.52$ & 2001-09-05 08:35:57 & 2001-09-05 11:22:35 & 10.9 & 7.9 & \phantom{0}2.8 \\
0093670301 & 257.97 & $-39.93$ & 2001-09-08 01:09:54 & 2001-09-08 04:17:28 & 14.8 & 15.1 & 10.9 \\
0093670401 & 258.87 & $-40.02$ & 2002-03-14 15:59:04 & 2002-03-14 19:37:57 & 11.4 & 11.7 & \phantom{0}6.3 \\
0093670501 & 258.37 & $-39.83$ & 2001-03-02 19:04:04 & 2001-03-02 21:37:24 & 12.6 & 12.7 & \phantom{0}7.5 \\
0203470401 & 258.55 & $-39.43$ & 2004-03-25 08:54:49 & 2004-03-25 12:36:08 & 15.6 & 15.9 & 10.3 \\
0203470501 & 258.00 & $-39.53$ & 2004-03-25 14:28:10 & 2004-03-25 18:09:30 & 13.7 & 13.7 & 10.9 \\
0207300201 & 258.37 & $-39.83$ & 2004-02-22 14:16:34 & 2004-02-22 23:42:50 & 13.1 & 13.7 & --- \\
0502080101 & 258.84 & $-39.66$ & 2007-09-15 05:12:23 & 2007-09-15 13:15:04 & 5.9 & 6.9 & 7.1 \\
0551030101 & 258.38 & $-40.20$ & 2008-09-27 17:46:20 & 2008-09-27 23:31:58 & 23.6 & 23.8 & 19.6 \\
0740830201 & 258.37 & $-39.83$ & 2014-03-02 07:43:38 & 2014-03-03 15:29:29 & 88.8 & 92.1 & --- \\
0743030101 & 258.93 & $-40.05$ & 2015-03-10 22:17:03 & 2015-03-11 20:47:16 & 66.3 & 67.2 & 54.1 \\
0804300101 & 257.91 & $-39.53$ & 2018-08-26 03:43:25 & 2018-08-27 03:17:50 & 82.7 & 83.0 & 68.3 \\
0804300301 & 258.00 & $-40.17$ & 2018-03-29 17:54:06 & 2018-03-30 14:55:31 & 62.5 & 64.2 & 45.7 \\
0804300401 & 258.44 & $-40.19$ & 2018-03-31 02:35:50 & 2018-04-01 00:06:36 & 65.8 & 72.6 & 38.2 \\
0804300501 & 258.35 & $-39.49$ & 2018-03-25 04:46:15 & 2018-03-26 15:12:56 & 103.8 & 108.0 & 84.2 \\
0804300601 & 258.99 & $-39.70$ & 2018-03-19 18:04:00 & 2018-03-20 15:40:49 & 60.3 & 62.3 & 53.7 \\
0804300701 & 258.80 & $-39.41$ & 2018-03-21 05:02:06 & 2018-03-22 03:24:32 & 76.0 & 81.1 & 60.1 \\
0804300801 & 258.69 & $-39.77$ & 2017-08-30 17:50:43 & 2017-08-31 05:39:09 & 44.6 & 44.6 & 38.8 \\
0804300901 & 258.34 & $-39.26$ & 2017-08-29 16:29:26 & 2017-08-30 04:04:54 & 27.8 & 28.6 & 19.7 \\
0804301001 & 257.92 & $-39.90$ & 2018-03-23 16:39:29 & 2018-03-24 15:18:32 & 20.8 & 67.7 & 48.8 \\
\enddata
\tablecomments{All exposure times represent the flare-filtered exposure.}
\label{tab1}
\end{deluxetable*}

The aim of the present paper is to revisit the comparison between the ISM and the gamma rays by using the new H.E.S.S. gamma-ray data in order to {pursue} the spatial correspondence revealed by \citetalias{2012ApJ...746...82F}. {This} paper is organized as follows. Section \ref{sec:obs} describes the datasets used in the work. Section \ref{sec:results} presents the results of the comparison among the gamma-rays, ISM, and X-rays. In Section \ref{sec:dis}, discussion on the comparison is presented and the origin of the gamma rays is explored into detail. Section \ref{sec:con} gives conclusions.

\section{Observational data} \label{sec:obs}
In Figures \ref{fig1}, we show the spatial distributions of the three datasets {rebinned to the} resolution of {the} H.E.S.S. data 1.4\,pc. They include the H.E.S.S. {$>$}2\,TeV energy band, the {\it XMM}-{\it Newton} X-rays, and the ISM, in Figures \ref{fig1}(a), \ref{fig1}(b), and \ref{fig1}(c), respectively, and Figure \ref{fig1}(d) shows the areas which indicate regions of different ISM column density. We explain their details in the following.

\subsection{X-Ray Data} \label{obs:Xray}
We used archival datasets obtained with {\it XMM}-{\it Newton} to create images of synchrotron X-rays in RX~J1713.7$-$3946. We utilized the {\it XMM}-{\it Newton} Science Analysis System \citep[SAS; ][]{2004ASPC..314..759G} version 18.0.0 and HEAsoft version 6.27.1 to analyze both the EPIC-pn and EPIC-MOS datasets with a total of 21 pointings (see Table \ref{tab1}). We reprocessed all the Observation Data Files (ODF) following standard procedures provided as the {\it XMM}-{\it Newton} extended source analysis software \citep[ESAS; ][]{2008A&A...478..575K}. The good time intervals (GTI) after filtering soft proton flares are shown in Table \ref{tab1}. To generate quiescent particle background (QPB) images and exposure maps for each observation, we run procedures of ``mos-/pn-back'' and ``mos-/pn-filter.'' The procedure ``merge\_comp\_xmm'' was used to combine the 21-pointing data. Note that we excluded 1) CCDs chips affected by strong stray light especially for high-energy bands and 2) short GTI data less than 1\,ks to improve the imaging quality. Finally, we applied an adaptive smoothing using the procedure ``adapt\_merge'', where the pixel sizes and smoothing counts were set to $6\arcsec$ and 80\,counts, respectively. We generated QPB-subtracted, exposure-corrected, and adaptively smoothed images in the energy bands of 0.5--0.8, 0.8--1.0, 1.0--2.0, 2.0--3.0, 3.0--5.0, and 1.0--5.0\,keV{\footnote{{Although the astrophysical X-ray background such as the Galactic diffuse X-ray emission is associated with the QPB-subtracted image, the background level is significantly lower than the synchrotron radiation from RX~J1713.7$-$3946 \citep[e.g.,][]{2018A+A...612A...6H}. It means that the small scale structures (or spatial variations of X-rays) of the SNR are not affected by the X-ray background. In the present paper, we did not subtract such X-ray background from the QPB-subtracted images.}}}. Because the X-ray image in the energy band of 1.0--5.0\,keV is not subject to significant absorption by the ISM (see Figures \ref{fig:app1} and \ref{fig:app2} in Appendix \ref{sec:app}), we used the 1.0--5.0\,keV\,image for {the} comparative analysis. {In the present analyses, we flagged two bright point sources, 1WGA~J1714.4$-$3945 and 1WGA~J1713.4$-$394, considering their point spread functions using the procedure ``cheese'' implemented in ESAS. We then rebinned the data to match grid size 4$\farcm$8 of the H.E.S.S. gamma-ray data. Here we averaged data values of X-rays within each pixel area, except for the point-sources flagged-areas. The rebinned X-ray image is presented as Figure \ref{fig1}(b).} The X-ray count $N_{\rm x}$ is given in unit of {$10^2$}\,photons\,s$^{-1}$\,degree$^{-2}$ and is used throughout the present paper.

\subsection{H.E.S.S. TeV Gamma-Ray Data} \label{obs:H.E.S.S.}
The gamma-ray data consist of two energy ranges, above 250\,GeV ({broad} energy band) and above 2\,TeV (high energy band), which are the combined data taken in several sessions over 2004, 2005, 2011, and 2012 \citep{2018A+A...612A...6H}.
The angular resolution  (full-width half maximum; FWHM) is 6\farcm 6 for the broad energy band and 4\farcm 8 for the high energy band, corresponding to the spatial resolution of 1.9\,pc and 1.4\,pc, respectively, at a distance of 1\,kpc.
{We also used the previous H.E.S.S. data in \cite{2007AA...464..235A}.The energy band is $>$300 GeV, similar to the broad energy band of \cite{2018A+A...612A...6H}, and the resolution is 8\farcm4 corresponding to 2.5 pc.}

{The $>$2 TeV dataset is named H.E.S.S.18($E>2$\,TeV), which provides the highest resolution and largest number of pixels among the H.E.S.S. data available. The $>$250 GeV dataset is named H.E.S.S.18($E>250$\,GeV) and the previous H.E.S.S data is named H.E.S.S.07. In addition to H.E.S.S.18($E>2$\,TeV) at $4\farcm8$ resolution, for testing dependence on angular resolution and gamma-ray energy, we analyze five H.E.S.S. datasets with lower resolution as listed in Table \ref{tab5}; two of them are H.E.S.S.18($E>2$\,TeV) smoothed to $6\farcm6$ resolution and H.E.S.S.18($E>250$\,GeV) having $6\farcm6$ resolution \citep{2018A+A...612A...6H}, and three of them are H.E.S.S.18($E>2$\,TeV) smoothed to $8\farcm4$ resolution, H.E.S.S.18($E>250$\,GeV) smoothed to $8\farcm4$ resolution, and  H.E.S.S.07 having $8\farcm4$ resolution. In the following analyses, all the X-ray data and the ISM data are binned to the same pixel size with the H.E.S.S. data.}
The gamma ray count {$N_{\rm{g}}$} {is given} in unit of excess counts\,arcmin$^{-2}$ throughout {this} paper. {The error was estimated as the square-root of the total excess counts within each pixel in conjunction with correction for the oversampling effect.}

The gamma-ray shell is peaked at $\sim$6\,pc in radius, beyond which the gamma rays show sharp decline. 
This suggests that {CR or electron} energy density decreases beyond the peak (\citealt{2018A+A...612A...6H}; \citetalias{2012ApJ...746...82F}). We assume that the CR energy density is nearly uniform within the shell and restrict the area of the analysis to the interior within the gamma-ray peak. By this restriction we eliminate the outer region where CR energy density decreases as shown by the white solid lines in Figure \ref{fig1}(a).

\begin{figure*}
\begin{center}
\includegraphics[width=160 mm,clip]{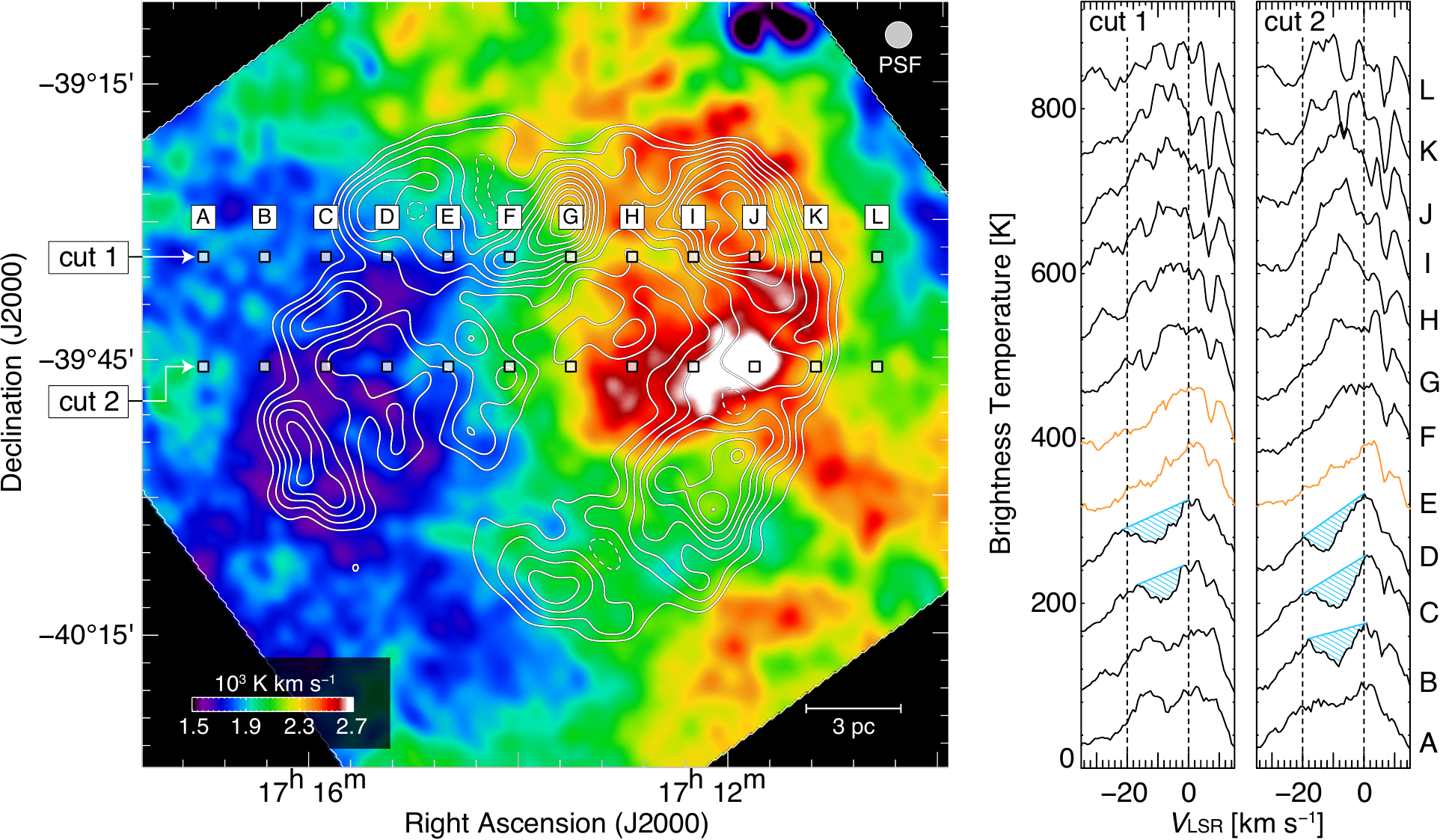}
\caption{Integrated intensity map of \ion{H}{1} overlaid with TeV gamma-ray contours ({\it{left panel}}), and \ion{H}{1} profiles ({\it{middle and right panels}}). The contour levels are the same as in Figure \ref{fig1}(d). The integration velocity range of \ion{H}{1} is from $-20.6$ to $-0.8$\,km\,s$^{-1}$. The \ion{H}{1} profiles are shown in along cuts 1 and 2. Each profile position (A--L) indicates square symbols labeled A--L in the left panel. The-self absorption dips are shown by the blue hatch with the straight background \ion{H}{1}, and the yellow profiles show the flat ones without correction for self absorption.}
\label{fig:HIabs}
\end{center}
\end{figure*}%

\subsection{ISM Proton Distribution} \label{obs:ISM}
We follow \citetalias{2012ApJ...746...82F} in calculating the ISM column density, which includes both molecular and atomic hydrogen. 
The molecular column density $N$(H$_2$) is calculated from the integrated intensity of the {$^{12}$CO($J=1$--0)} emission $W$(CO) (K\,km\,s$^{-1}$) for an $X_{\rm CO}$ factor $2\times10^{20}$\,cm$^{-2}$ (K\,km\,s$^{-1})^{-1}$ \citep{1993ApJ...416..587B} {which is typically found in the Galaxy};
\begin{eqnarray}
N_{\mathrm p}({\mathrm H}_2)\;(\mathrm{cm}^{-2}) & = & 2 \times N({\mathrm H}_2) \nonumber \\   
& = & 4 \times 10^{20}\;W({\rm CO})\;\mbox{(K\,km\,s$^{-1}$)}.
\label{equ:1}
\end{eqnarray}
The atomic hydrogen column density $N$(\ion{H}{1}) (cm$^{-2}$) was calculated from the integrated intensity $W$(\ion{H}{1}) (K\,km\,s$^{-1}$) of the 21\,cm \ion{H}{1} emission by adopting the optically thin approximation as follows;
\begin{equation}
N(\mbox{\ion{H}{1}}) \;(\mathrm{cm}^{-2})  = 1.832 \times 10^{18}\; W(\mbox{\ion{H}{1}})\; \mbox{(K\,km\,s$^{-1}$)}. 
\label{equ:2}
\end{equation}  
{In the present paper, we used the velocity integration range of $W$(CO) or $W$(\ion{H}{1}) from $-20$ to $0$ km s$^{-1}$, which is compatible with the previous study \citepalias[][]{2012ApJ...746...82F}.}

{Figure \ref{fig1}(c)shows the ISM column density distribution in units of 10$^{22}$ cm$^{-2}$. We averaged values of column density within each pixel area, as with the X-ray datasets.} We also make correction for the \ion{H}{1} self-absorption in the SE cold \ion{H}{1} cloud by using the method in \citet{1978AJ.....83.1607S}, where the background \ion{H}{1} emission is assumed by a straight line as explained in \citetalias{2012ApJ...746...82F}. Some of the other positions may be subject to self-absorption. Figure \ref{fig:HIabs} shows two cuts of \ion{H}{1} profiles toward the region including \ion{H}{1} with self-absorption, and indicates that a few spectra have clear dips and that some have flat spectra have clear dips in addition to some having flat spectra, e.g., E and F in cut1 and cut2. It is possible that the flat spectra are due to self-absorption, whereas there is no reliable way to assume background \ion{H}{1} without self-absorption. If we assume a background \ion{H}{1} emission, identical to that at position H in cut2, we estimate $W$(\ion{H}{1}) becomes larger than the optically thin case by a factor of 2, while the correction is arbitrary. We therefore selected {18} pixels by eye inspection as those with uncertain background emission as enclosed by the gray solid lines in Figure \ref{fig1}(c), and excluded them in the present analysis in order not to impair the accuracy. {It is possible that the \ion{H}{1} emission is affected by the optically thick \ion{H}{1} \citep{2014ApJ...796...59F, 2015ApJ...798....6F, 2018ApJ...860...33F}, whereas the direction of RX~J1713 is in the Galactic plane, not allowing to make a direct estimate of the \ion{H}{1} optical depth by using the Planck dust emission \citep{2014A&A...571A..11P}. We therefore did not attempt to apply further the \ion{H}{1} optical depth correction.}

The total hydrogen column density is calculated as follows;
\begin{equation}
N_\mathrm{p} = N_\mathrm{p}({\mathrm H}_2+\mbox{\ion{H}{1}}) = 2 \times N_\mathrm{p}({\mathrm H}_2)\; +\; N(\mbox{\ion{H}{1}}).
\label{equ:3}
\end{equation}  
Considering these {ISM proton} properties, we denoted the two regions as shown in Figure \ref{fig1}(c); the region where self-absorption is corrected for (14 pixels shown by the red solid lines) is included in the analysis, and that where correction for self-absorption is difficult due to a flat line shape, {18} pixels shown by the gray solid lines, are masked in the analysis. 
{According to F12, the errors in the region with self-absorption are estimated to be $<$20\%, while the errors elsewhere are mostly statistical ones and negligibly small.}
In Figure \ref{fig1}(d) we divide the pixels into five zones {of $N_{\rm p}$}, which are used in Section \ref{sec:results}.
The total number of the pixels after this masking is {75} in H.E.S.S.18 ($E>2$\,TeV).
A similar masking was also applied to H.E.S.S.07 and the total number of the pixels is 32.

Details of the respective ISM observations are summarized below.

\begin{figure*}
\begin{center}
\includegraphics[width=\linewidth,clip]{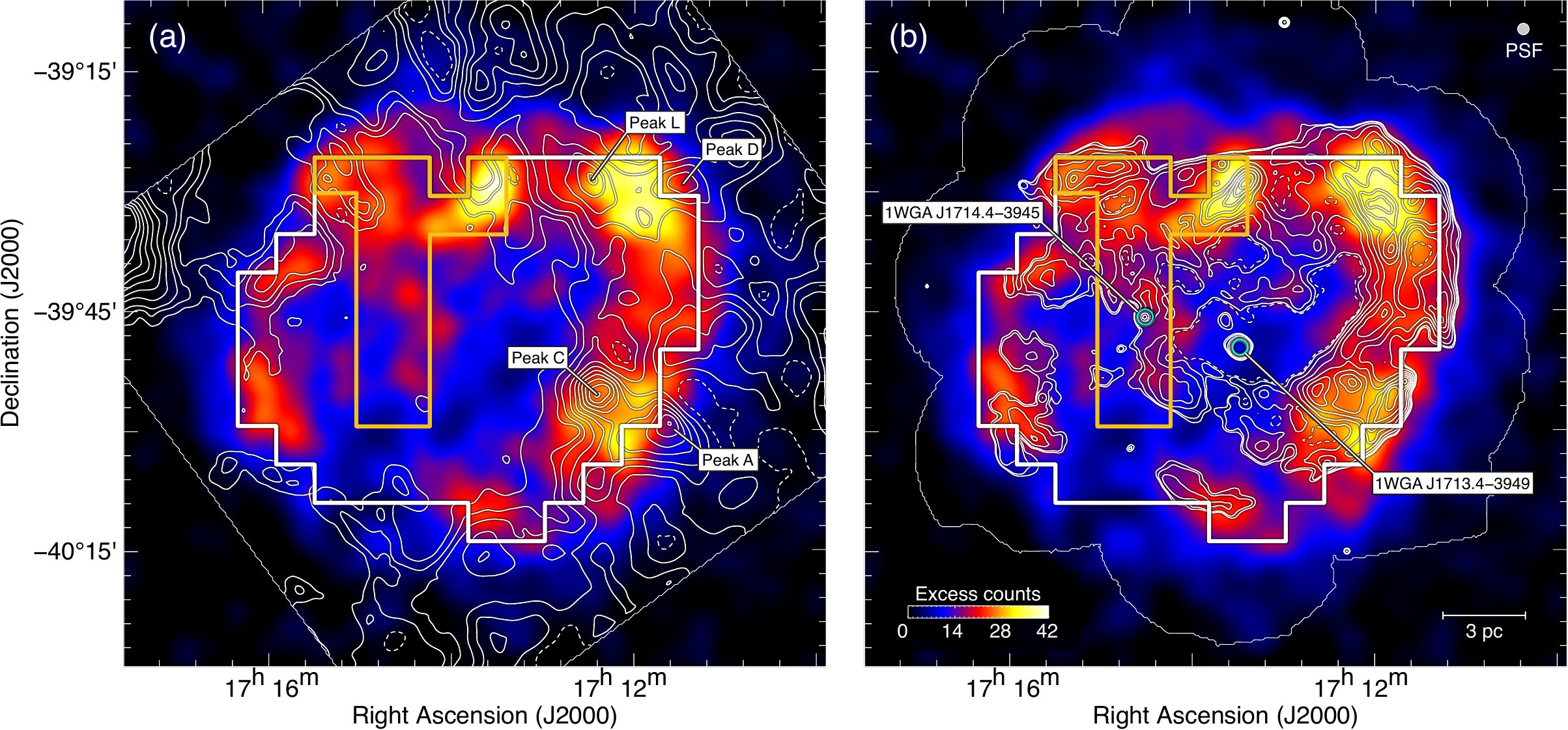}
\caption{Maps of TeV gamma-rays {($E > 2$\,TeV)} toward RX~J1713.7$-$3946 overlaid with (a) total proton column density $N_\mathrm{p}$(H$_2+$\ion{H}{1}) \citepalias{2012ApJ...746...82F} and (b) {\it{XMM-Newton}} X-rays {($E$: 1.0--5.0\,keV)}. The contour levels are 6, 7, 8, 10, 12, 14, 16, 18, 20, and $22 \times 10^{21}$\,cm$^{-2}$ for $N_\mathrm{p}$(H$_2+$\ion{H}{1}); and {37, 40, 49, 64, 85, 112, 145, and 184\,photons\,s$^{-1}$\,degree$^2$} for X-rays. The white polygon is the same as shown in Figure \ref{fig1}. {The orange polygon is the same as gray polygon in Figure \ref{fig1}(c).} Thin rectangle and curve indicate observed regions of $N_\mathrm{p}$(H$_2+$\ion{H}{1}) and X-rays, respectively. The positions of CO peaks A, C, D, and L are shown. The two prominent X-ray point sources, 1WGA~J1714.4$-$3945 and 1WGA~J1713.4$-$3949, are also shown in circles.} 
\label{fig:whole}
\end{center}
\end{figure*}%

\subsubsection{The NANTEN CO Data} \label{ISM:NANTEN}
We utilize the $^{12}$CO($J = 1$--0) data at 115.271202\,GHz obtained with the NANTEN 4-m millimeter telescope, which were published by \citet{2005ApJ...631..947M}. The observations were carried out by using the position-switching method with a grid spacing of 2\farcm 0. The mapped area is $\sim$1.9 degree$^2$ in the region of 346\fdg 7--348\fdg 0 in Galactic longitude and of $-1\fdg 2$--0\fdg 2 in Galactic latitude, including CO peaks A, C, D and L (\citetalias{2003PASJ...55L..61F}; \citealt{2005ApJ...631..947M, 2013ApJ...778...59S}). The 4\,K cooled Nb superconductor-insulator-superconductor (SIS) mixer was used as the frontend \citep{1990IJIMW..11..717O}. The typical system temperature was $\sim$250\,K in the single-sideband (SSB), including the atmosphere toward the zenith. The backend was an acousto-optical spectrometer (AOS) with 2048 channels and 250\,MHz bandwidth. The CO data cover a velocity range of 600\,km s$^{-1}$ with a velocity resolution of 0.65\,km\,s$^{-1}$. The final data have a beam size of 156 arcsec and typical noise fluctuations of 0.2--0.3\,K\,ch$^{-1}$ \citep[more detailed information, see Section 2 of][]{2005ApJ...631..947M}. 

\subsubsection{The ATCA plus Parkes \ion{H}{1} Data} \label{ISM:ATCA}
The \ion{H}{1} data at\,1.4 GHz are from the Southern Galactic Plane Survey \citep{2005ApJS..158..178M}, which were obtained by combining the \ion{H}{1} datasets with the Australia Telescope Compact Array (ATCA) and the Parkes Radio Telescope. The beam size $2\farcm2$ is high enough to spatially resolve the cool \ion{H}{1} cloud in the southeast-rim with {a} self-absorption profile. The typical noise fluctuations are 1.9\,K at a velocity resolution of 0.82\,km\,s$^{-1}$.

\section{Correlation among the gamma rays, the ISM and the X-rays} \label{sec:results}
Figures \ref{fig:whole}(a) and \ref{fig:whole}(b) show gamma-ray distribution superposed on the ISM contours and X-ray contours, respectively, which suggest that the three distributions are fairly similar {to} each other. This is consistent with the hybrid picture of the two gamma-ray origins (Section \ref{sec:intro}).
{In the following, we describe the analyses of H.E.S.S.18($E>2$\,TeV) in Sections \ref{res:3D_IXG} and \ref{res:ha_le}, and those of H.E.S.S.18($E>250$\,GeV) and H.E.S.S.07, and of the spatially smoothed H.E.S.S.18 data in Section \ref{subsec:low_reso}.}

\subsection{A Unified Fitting in the 3D Space of $N_{\rm p}$--$N_{\rm x}$--$N_{\rm g}$} \label{res:3D_IXG}

\begin{deluxetable*}{lCRRRRRRC}
\tablecaption{Summary of the multiple linear regression}\label{tab:summary_multireg}
\tablewidth{0pt}
\label{tab5}
\tablehead{
\multicolumn4c{Dataset} & & & & \multicolumn2c{$F$-test} \\
\cline{1-4}
\cline{8-9}
\colhead{Energy band} & \colhead{Pixel size} & \colhead{$n$} & \colhead{{red}{$r$}} & \colhead{$a$} & \colhead{$b$} & \colhead{$R^{2}$} & \colhead{$F$} & \colhead{$F_{0.01}(2, n-2)$}}
\decimalcolnumbers
\startdata
\multicolumn{9}{l}{H.E.S.S.18} \\
\cline{1-9}
$E>2$\,TeV & 4\farcm 8 & 75 & 0.71 & 1.45\pm 0.17 & 1.02\pm 0.24 & 0.92 & 441.1 & 4.9  \\
 & 6\farcm 6 & 36 & 0.74 & 1.39\pm 0.24 & 1.10\pm 0.34 & 0.97 & 660.0 & 5.3  \\
 & 8\farcm 4 & 23 & 0.80 & 1.37\pm 0.28 & 1.05\pm 0.41 & 0.98 & 646.9 & 5.8 \\
$E>250$\,GeV & 6\farcm 6 & 36 & 0.74 & 10.6\phn\pm 1.7\phn & 7.9\phn\pm 2.4\phn & 0.97 & 641.3 & 5.3 \\
 & 8\farcm 4 & 23 & 0.80 & 10.4\phn\pm 1.8\phn & 8.1\phn\pm 2.6\phn & 0.99 & 896.1 & 5.8 \\
\cline{1-9}
\multicolumn{9}{l}{H.E.S.S.07} \\
\cline{1-9}
$E \gtrsim 300$\,GeV\tablenotemark{a} & 8\farcm 4 & 23 & 0.80 & 1.87\pm 0.27 & 1.07\pm 0.40 & 0.97 & 383.0 & 5.8
\enddata
\tablenotetext{a}{{We presumed the same energy threshold that was used for the spectral analysis in \cite{2007AA...464..235A}.}}
\tablecomments{
Columns (1) -- (3): energy band, pixel size and number of pixels of the dataset, (see Section \ref{obs:H.E.S.S.}), {red}{(4) correlation coefficient between $N_\mathrm{p}$ and $N_\mathrm{x}$}, (5) and (6): regression coefficients estimated by Equation (\ref{eqn:vector_estimated_b}) and standard deviations of them (the square root of the diagonal elements of Equation (\ref{eqn:standard_error_matrix})). (7): coefficient of determination given by Equation (\ref{eqn:R2}), (8): $F$-statistic given by Equation (\ref{eqn:f-statistic}) and (9) critical $F$-value with $(2, n-2)$ degrees of freedom for 1\% significance level.
See Appendix \ref{sec:multiple_regression} for further details.}
\vspace*{-0.2cm}
\end{deluxetable*}

Motivated by the basic picture above regarding the gamma-ray origin, we formulate the gamma-ray counts as a combination of two terms, one proportional to $N_{\rm p}$ and the other proportional to $N_{\rm x}$ as a first approximation. Simple approximate relationships of $N_{\rm g}$ are given in a line of sight as follows; the hadronic gamma rays $N_{\rm g}$(hadronic) via the p-p reaction is proportional to the CR proton column density $N_{\rm p}$(CR) times the ISM proton column density $N_{\rm p}$(ISM) [$N_{\rm g}$(hadronic) $\propto$ $N_{\rm p}$(CR)$ \times N_{\rm p}$], and the leptonic gamma rays via the inverse Compton scattering $N_{\rm g}$(leptonic) is proportional to the CR electron column density $N_{\rm e}$(CR) times the uniform low-energy photon density $N_{\rm photon}$(CMB$=${the} Cosmic Microwave Background) [$N_{\rm g}$(leptonic) $\propto$ $N_{\rm e}$(CR)  $ \times N_{\rm photon}$(CMB)]. Noting that the nonthermal X-ray count $N_{\rm x}$ is proportional to $N_{\rm e}$(CR) times the squared magnetic field flux density $B^2$ [$N_{\rm x} \propto N_{\rm e}$(CR) $\times B^2$], $N_{\rm g}$(leptonic) is hence proportional to $N_{\rm x}$ times ($N_{\rm photon}$(CMB) $/ B^2$) [$N_{\rm g}$(leptonic) $ \propto N_{\rm x} \times$ ($N_{\rm photon}$(CMB)$ /B^2$)] \citep[e.g.,][]{1997MNRAS.291..162A}. 

As a result, $N_\mathrm{g}$ for the combination is given as follows;
\begin{equation}\label{equ:4}
N_\mathrm{g}=aN_\mathrm{p}+bN_\mathrm{x},
\end{equation}
where the coefficient $a$ is proportional to $N_\mathrm{p}\mbox{(CR)}$ and coefficient $b$ to $N_\mathrm{photon}\mbox{(CMB)}/B^{2}$.
{Here, we can see that the $a$ and $b$ coefficients respectively encapsulate a global dependence on the {cosmic ray energy} density and the CMB/$B$-field energy density ratio, as a way to provide estimates of the hadronic and leptonic fractions.} {We assume here that the magnetic field strength $B$ is uniform in most of the volume as suggested by \citetalias{2012ApJ...744...71I}. We consider that the CMB photon density is also uniform and that additional stellar photons are insignificant {\citep[e.g.,][]{2008ApJ...685..988T}}. In addition, the energy density of CRs is assumed to be uniform within the SNR shell which is consistent with the model calculations (e.g., \citetalias[e.g.,][]{2010ApJ...708..965Z}).}

\begin{figure*}
\begin{center}
\includegraphics[]{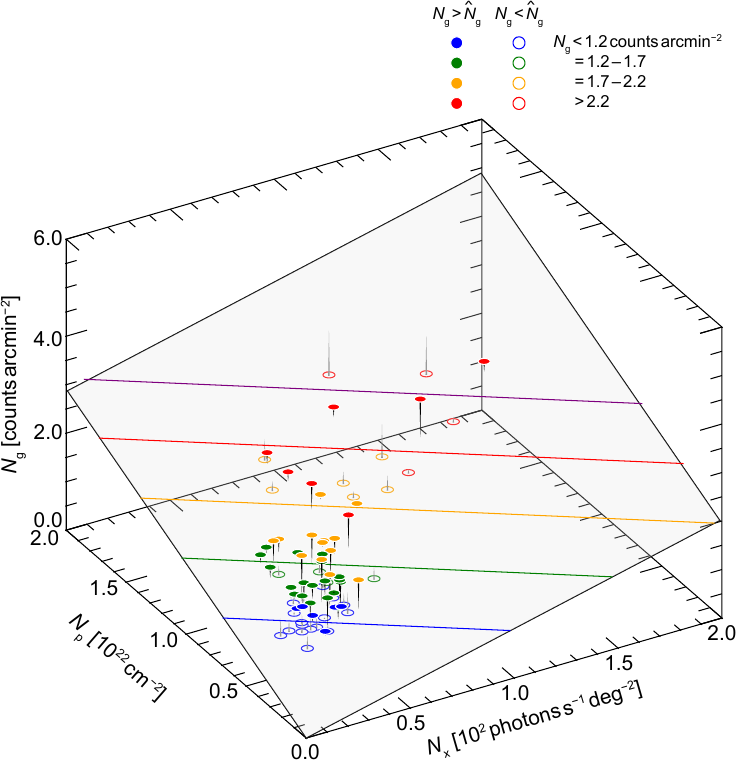}
\caption{3D fitting of a flat plane expressed by Equation (\ref{equ:5}) in the $N_\mathrm{p}$-$N_\mathrm{x}$-$N_\mathrm{g}$ space for H.E.S.S.18 ($E>2$\,TeV) with a pixel size of $4\farcm 8$. 
The data pixels are colored by the code in the figure according to $N_{\rm g}$, which are shown by the filled and open symbols for those above and below the plane.
Each vertical (parallel to the $N_\mathrm{g}$ axis) line connects $N_\mathrm{g}$ and $\hat{N_\mathrm{g}}$. {The blue, green, orange, red, and purple lines on the best-fit plane indicate $\hat{N_\mathrm{g}}=1.0$, 1.5, 2.0, 2.5, and 3.0 counts arcmin$^{-2}$, respectively.}
}
\label{fig:3dpl}
\end{center}
\end{figure*}

\begin{figure}
\begin{center}
\includegraphics[width=\linewidth,clip]{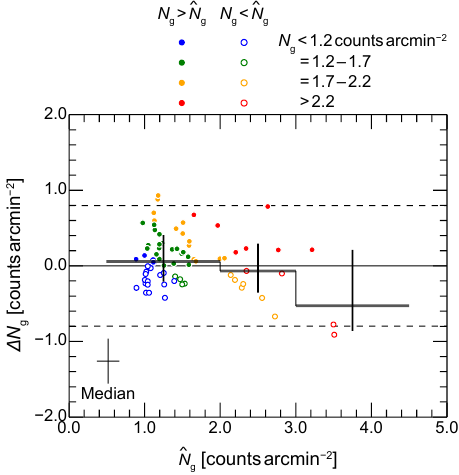}
\caption{
A plot of the difference $\Delta N_\mathrm{g}=N_\mathrm{g}-\hat{N}_\mathrm{g}$ with respect to $\hat{N}_{\rm g}$ derived from Equation (\ref{equ:4}) for H.E.S.S.18 ($E>2$\,TeV) with a pixel size of $4\farcm 8$.
The positive (above the horizontal dashed line) pixels and the negative pixels are shown by filled and open symbols, and the color code in $N_{\rm g}$ is the same as in Figure \ref{fig:3dpl}.
The horizontal- and vertical error bars shown in the left bottom corner indicate the median values of $\sigma(\hat{N_\mathrm{g}})$ (Equation (\ref{eqn:sigma_hat_ngi})) and $\sigma(\Delta N_\mathrm{g})$ (Equation (\ref{eqn:sigma_delta_ngi})), respectively.
The averages of $\Delta N_\mathrm{g}$ {weighted with $\sigma(\Delta N_\mathrm{g})^{-2}$} are shown for three bins of $\hat{N}_\mathrm{g}$ with the vertical error bars.
{The dashed lines show the $2.5\times$\,weighted-root-mean-square of $\Delta N_\mathrm{g}$.}
}
\label{fig:3derr}
\end{center}
\end{figure}

We apply Equation (\ref{equ:4}) to the data pixels in a 3D space of $N_\mathrm{p}$-$N_\mathrm{x}$-$N_\mathrm{g}$.
For this purpose, we developed the formulation for the least-squares fitting as given in Appendix \ref{subsec:multiple_regression_coeff} in the present work {and applied it to H.E.S.S.18($E>2$\,TeV)}.
As a result, we obtained a multiple linear regression plane given by the following relation;
\begin{equation}\label{equ:5}
\hat{N_\mathrm{g}}=(1.45\pm 0.17)\times N_\mathrm{p}+(1.02\pm 0.24)\times N_\mathrm{x},
\end{equation}
for all the {75} pixels across the ISM/X-ray/gamma 3D space.
The hat symbol (\,$\hat{}$\,) put on $N_\mathrm{g}$ means that it is predicted by the regression.
Table \ref{tab5} {lists $a$ and $b$, and} shows that the fitting is reasonably good with the coefficient of determination $R^{2}=0.92$ (see Appendix \ref{subsec:goodnes_of_fit}), which means that 90\% of the variance in $N_\mathrm{g}$ is predictable from $N_\mathrm{p}$ and $N_\mathrm{x}$.
Figure \ref{fig:3dpl} shows the plane with the {75} pixels, and we confirm the good fitting.
In Figure \ref{fig:3derr} we show a plot of $\Delta N_\mathrm{g}=N_\mathrm{g}-\hat{N_\mathrm{g}}$ as a function of $\hat{N_\mathrm{g}}$.
This corresponds to a side view of the plane in Figure \ref{fig:3dpl} from the direction of ${\hat{N_\mathrm{g}}}=\mbox{constant}$.
$\Delta N_\mathrm{g}$ is nearly zero at $\hat{N_\mathrm{g}}$ below {2.0}, showing a good fit of the regression for most of the pixels.
There seems to be a hint of slight decrease of $\Delta N_\mathrm{g}$ at $\hat{N_\mathrm{g}}$ above {2.0}, although the number of the pixels in this category is {15}.
The uncertainty in $\hat{N_\mathrm{g}}$ for $i$-th pixel {($i=1 \cdots 75$)} is given by
\begin{eqnarray} \label{eqn:sigma_hat_ngi}
\sigma(\hat{N}_{\mathrm{g},\:i}) & = & \left\{(N_{\mathrm{p},\:i}\,\sigma_{a})^{2}+[a\,\sigma(N_{\mathrm{p},\: i})]^{2}\right. \nonumber \\
 & & \left.+(N_{\mathrm{x},\:i}\,\sigma_{b})^{2}+[b\,\sigma(N_{\mathrm{x},\:i})]^{2}\right\}^{1/2},
\end{eqnarray}
and that of $\Delta N_\mathrm{g}$ by
\begin{equation} \label{eqn:sigma_delta_ngi}
\sigma(\Delta N_{\mathrm{g},\:i})=\left[{\sigma(N_{\mathrm{g},\:i}})^{2}+{\sigma(\hat{N}_{\mathrm{g},\:i}})^{2}\right]^{1/2}.
\end{equation}
Here, $\sigma(N_\mathrm{g})$, $\sigma(N_\mathrm{p})$ and $\sigma(N_\mathrm{x})$ are the uncertainties in $N_\mathrm{g}$, $N_\mathrm{p}$ and $N_\mathrm{x}$, {and} $\sigma_{a}$ and $\sigma_{b}$ are the standard errors of the estimated $a$ and $b$ (Equation (\ref{eqn:standard_error_matrix}) in Appendix \ref{subsec:multiple_regression_coeff}), respectively.
Note that we have estimated $\sigma(N_{\mathrm{p},\:i})$ from the rms noise in $W(\mbox{\ion{H}{1}})$ and $W(\mbox{CO})$, without taking into account uncertainty in the $X_\mathrm{CO}$ factor.

\begin{deluxetable*}{lCRRRRRRR}
\tablecaption{Estimate of the hadronic- and leptonic-origin gamma-rays} \label{tab:estimate_hadron_and_lepton}
\tablewidth{0pt}
\label{label}
\tablehead{
\multicolumn2c{Dataset} & & & \multicolumn2c{Hadronic component} & & \multicolumn2c{Leptonic component} \\
\cline{1-2}
\cline{5-6}
\cline{8-9}
\colhead{Energy band} & \colhead{Pixel size} & \colhead{$\langle\hat{N_\mathrm{g}}\rangle$} & \colhead{$\langle N_\mathrm{p}\rangle$} & \colhead{$\hat{N}_\mathrm{g}^\mathrm{hadronic}$} & \colhead{$\hat{N}_\mathrm{g}^\mathrm{hadronic}/\langle\hat{N_\mathrm{g}}\rangle$}  & \colhead{$\langle N_\mathrm{x}\rangle$} & \colhead{$\hat{N}_\mathrm{g}^\mathrm{leptonic}$} & \colhead{$\hat{N}_\mathrm{g}^\mathrm{leptonic}/\langle \hat{N_\mathrm{g}}\rangle$}
}
\decimalcolnumbers
\startdata
\multicolumn{9}{l}{H.E.S.S.18} \\
\cline{1-9}
$E>2$\,TeV & 4\farcm 8 & 1.56\pm 0.02 & 0.72 & 1.04\pm 0.12 & (67\pm\phn 8)\% & 0.50 & 0.51\pm 0.12 & (33\pm\phn 8)\% \\
 & 6\farcm 6 & 1.57\pm 0.04 & 0.73 & 1.01\pm 0.17 & (65\pm 11)\% & 0.50 & 0.55\pm 0.17 & (35\pm 11)\% \\
 & 8\farcm 4 & 1.54\pm 0.07 & 0.74 & 1.01\pm 0.21 & (66\pm 14)\% & 0.50 & 0.53\pm 0.20 & (34\pm 13)\% \\
$E>250$\,GeV & 6\farcm 6 & 11.7\phn\pm 0.3\phn & 0.73 & 7.7\phn\pm 1.2\phn & (66\pm 11)\% & 0.50 & 4.0\phn\pm 1.2\phn & (34\pm 10)\% \\
 & 8\farcm 4 & 11.7\phn\pm 0.4\phn & 0.74 & 7.6\phn\pm 1.3\phn & (65\pm 11)\% & 0.50 & 4.0\phn\pm 1.3\phn & (35\pm 11)\% \\
\cline{1-9}
\multicolumn{9}{l}{H.E.S.S.07} \\
\cline{1-9}
$E\gtrsim 300$\,GeV & 8\farcm 4 & 1.92\pm 0.06 & 0.74 & 1.38\pm 0.20 & (72\pm 11)\% & 0.50 & 0.54\pm 0.20 & (28\pm 10)\% \\
\enddata
\tablecomments{
Columns (1) and (2): energy band and pixel size of the dataset, (3), (4) and (7): spatial averages of observed $N_\mathrm{g}$ (counts\,arcmin$^{-2}$), $N_\mathrm{p}$ ($10^{22}$\,cm$^{-2}$) and $N_\mathrm{x}$ ($10^2$ photons\,s$^{-1}$\,degree$^{-2}$), (5) and (8): predicted values of hadronic- and leptonic-origin gamma-rays (counts\,arcmin$^{-2}$), (6) and (9): fraction of the hadronic and leptonic components. 
}
\end{deluxetable*}

\subsection{Separation of the Hadronic and Leptonic Origins} \label{res:ha_le}
Equation (\ref{equ:5}) allows us to make a first estimate of the gamma rays of the hadronic and leptonic origins separately.
The hadronic component is given by
\begin{equation}
\hat{N}_\mathrm{g}^\mathrm{hadronic} = aN_\mathrm{p}
\end{equation}
and its uncertainty is
\begin{eqnarray}
\sigma(\hat{N}_\mathrm{g}^\mathrm{hadronic}) & = & \left\{(\sigma_{a}\,N_\mathrm{p})^{2}+[a\,\sigma(N_\mathrm{p})]^{2}\right\}^{1/2} \nonumber \\
 & \sim & \sigma_{a}\,N_\mathrm{p},
\end{eqnarray}
where we ignore the second term because $\sigma(N_\mathrm{p})/N_\mathrm{p}=10^{-2}$ is one-order of magnitude smaller than $\sigma_{a}/a$.
The leptonic component and its uncertainty are given in a similar manner,
\begin{equation}
\hat{N}_\mathrm{g}^\mathrm{leptonic} = bN_\mathrm{x}
\end{equation}
and
\begin{equation}
\sigma(\hat{N}_\mathrm{g}^\mathrm{leptonic}) \sim \sigma_{b}\,N_\mathrm{x},
\end{equation}
respectively.
By using the coefficients $a$ and $b$ obtained in Section \ref{res:3D_IXG} (summarized in Table \ref{tab:summary_multireg}) and the spatially-averaged gamma-ray count $\bar{N_\mathrm{g}}=\sum_{i}N_{\mathrm{g},\:i}/n$, ISM proton column density $\bar{N_\mathrm{p}}=\sum_{i}N_{\mathrm{p},\:i}/n$ and X-ray count $\bar{N_\mathrm{x}}=\sum_{i}N_{\mathrm{x},\:i}/n$ listed in Table \ref{tab:estimate_hadron_and_lepton}, it is estimated that the hadronic component occupies {$(67\pm 8)$\%} of the total gamma ray count and the leptonic component {$(33\pm 8)$\%}.
Further discussion is given in Section \ref{subsec:hybridorigin_in_SNR}.

{We checked the effect of adding back the excluded 18 pixels and found that the fitted values for $a = 1.20 \pm 0.18$ and $b = 1.52 \pm 0.24$ (see Equation \ref{equ:4}) suggested hadronic and leptonic fractions of 52$\pm$8\% and 48$\pm$8\% respectively. The hadronic fraction has decreased compared to our main result, as expected, due to the lack of self-absorption corrected leads to an underestimate of $N_\mathrm{p}$.}


\subsection{{Positional dependence of the hadronic gamma-ray fraction}} \label{subsec:had_frac}

\begin{figure}
\begin{center}
\includegraphics[width=\linewidth,clip]{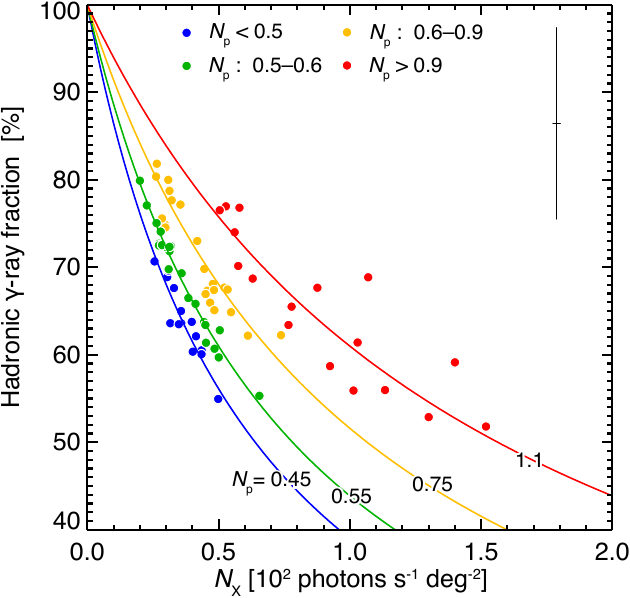}
\caption{{Scatter plot between hadronic gamma-ray fraction and $N_{\rm{x}}$. The hadronic gamma-ray fraction is derived from Equation (\ref{equ:a13}). Blue, green, yellow, and red circles represent the data points where $N_{\rm p}$ less than 0.5, 0.5--0.6, 0.6--0.9, and more than 0.9, respectively. Blue, green, yellow, and red solid lines show the relationships derived from Equation (\ref{equ:a13}) when $N_{\rm p}=0.45, 0.55, 0.75,$ and 1.1, respectively. The typical error bar for each data point is shown in the right top corner, which indicates median values of $\sigma(N_{\rm p})$ and hadronic gamma-ray fraction. The error of the hadronic gamma-ray fraction is estimated as $(aN_{\rm p}/\hat{N}_\mathrm{g})((\sigma_a/a)^2+(N_{\rm p}/\sigma(N_{\rm p}))^2+(\hat{N}_\mathrm{g}/\sigma(\hat{N}_\mathrm{g}))^2)^{1/2}$}.}
\label{fig:fracplot}
\end{center}
\end{figure}

{We have obtained the hadronic and leptonic gamma-ray count in each pixel, and are able to compare [the fraction of the hadronic$/$leptonic gamma rays in the total gamma rays] and [$N_{\rm p}$ or $N_{\rm x}$] over the SNR. In Figure \ref{fig:fracplot}, we present a scatter plot between the hadronic gamma ray fraction and $N_{\rm x}$ in all 75 pixels for different ranges of $N_\mathrm{p}$ values. The hadronic fraction fitted by the model is expressed from Equation (\ref{equ:4}) as follows;}

\begin{figure*}
\begin{center}
\includegraphics[width=\linewidth,clip]{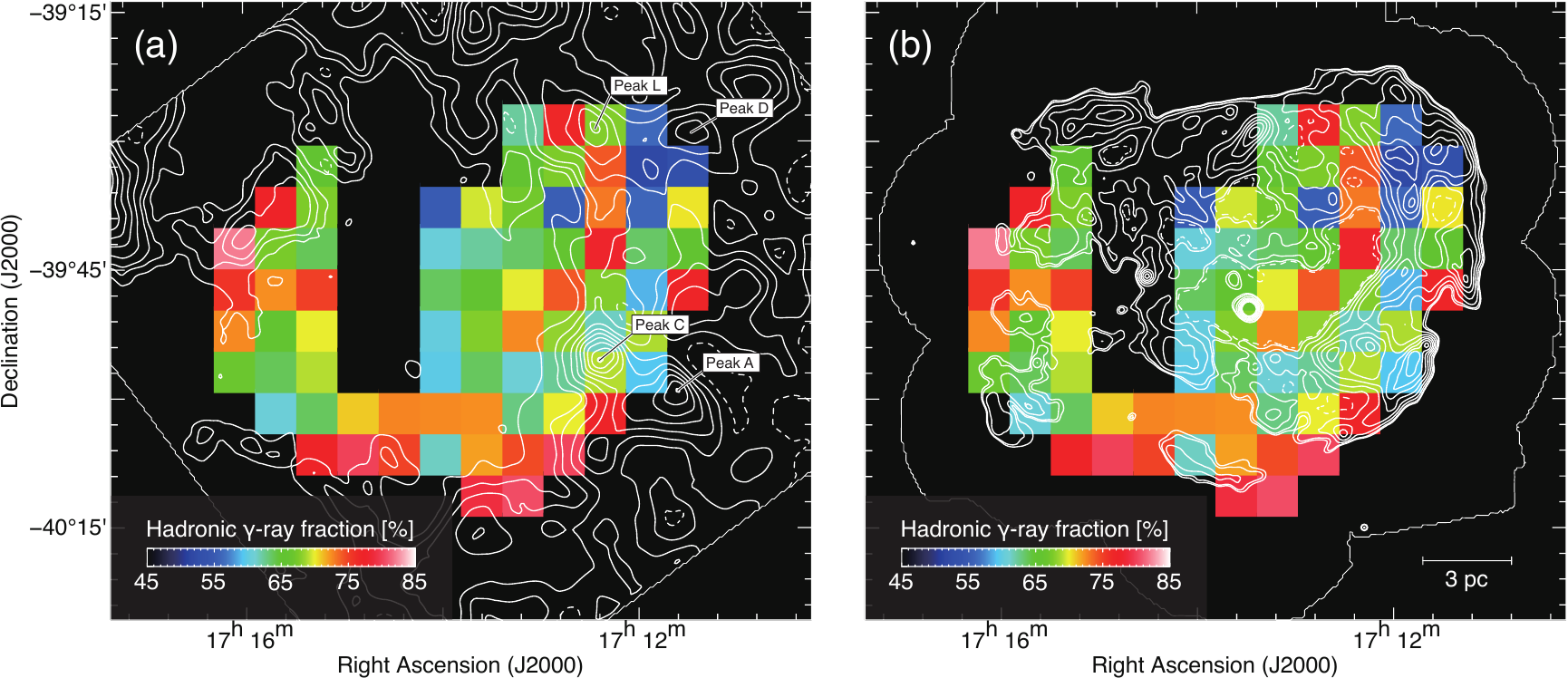}
\caption{{Distributions of hadronic gamma-ray fraction overlaid with (a) total proton column density $N_\mathrm{p}$(H$_2+$\ion{H}{1}) (\citetalias{2012ApJ...746...82F}) and (b) {\it{XMM-Newton}} X-rays {($E$: 1.0--5.0\,keV)}. The contour levels are the same as shown in Figure \ref{fig:whole}. The prominent dense clumps Peak A, C, D, and L are also indicated. }}
\label{fig:fracmap}
\end{center}
\end{figure*}

{
\begin{equation}
\mathrm{[Hadronic\;gamma\;ray\;fraction]}=\frac{aN_\mathrm{p}}{\hat{N}_\mathrm{g}},
\label{equ:a13}
\end{equation} 
which indicates that the fraction increases with $N_{\rm p}$ and decreases with $N_{\rm x}$. The relationships derived from Equation (\ref{equ:a13}) are plotted by the colored thin solid lines in Figure \ref{fig:fracplot}, showing a good correspondence with each pixel. The model includes both the hadronic and leptonic components and Figure \ref{fig:fracplot} naturally shows a mixed trend. 
We find that a high hadronic fraction more than 70 $\%$ is found toward low $N_{\rm x} < 0.6$ but not only toward high $N_{\rm p} > 0.9$. It is also notable that more than half of the pixels with a low hadronic fraction less than 60 $\%$ have high $N_{\rm p} > 0.9$. These trends are likely caused by the enhanced X-rays toward dense ISM clumps. {Figure \ref{fig:fracplot} indicates that there are hadronic-dominated regions with $>70$\%, whereas there are no such regions dominated by leptonic emission to the same level.} We also recognize these trends in Figures \ref{fig:fracmap}(a) and \ref{fig:fracmap}(b), which show overlays of the hadronic fraction with the distributions of $N_{\rm p}$ and $N_{\rm x}$, respectively. The two figures are somewhat complicated to grasp simply, whereas we recognize that a high hadronic fraction is often found toward regions of low $N_{\rm x}$ as is consistent with Figure \ref{fig:fracplot}; they are in the south-eastern rim and in the east of the western bright X-ray shell. These comparisons will provide observational signatures of the two gamma-ray origins, which will be revealed more into depth at higher resolution in the CTA era.
}

\subsection{{Analyses at lower resolutions}} \label{subsec:low_reso}
{We made additional analyses of the five H.E.S.S. datasets at resolutions of 6$\farcm$6 and 8$\farcm$4 as listed in Table \ref{tab5}. The method explained in Sections \ref{res:3D_IXG} is applied to them and regression flat planes derived are included in Table \ref{tab5}. The planes are used to derive the hadronic and leptonic gamma rays in the same manner explained in Section \ref{res:ha_le}.}

{In Table \ref{label} we find that the hadronic and leptonic gamma rays show similar fractions of (65--72)\%:(28--35)\% {or $\sim$6$\sigma$:3$\sigma$} for each resolution. The overall results are consistent with each other and do not depend significantly on the resolution and energy. The H.E.S.S.18($E>2$\,TeV) results yield the hadronic and leptonic counts of $\sim$8$\sigma$:4$\sigma$ at the highest resolution, and are of somewhat higher relative sensitivity than the lower resolution results.}

\section{Discussion on the gamma-ray origin} \label{sec:dis}
\subsection{The Hybrid Origin of the Gamma Rays in a SNR} \label{subsec:hybridorigin_in_SNR}
The present analysis revealed that the gamma rays in RX~J1713 consist of the hadronic components and the leptonic components with a ratio of {$7:3$} in the gamma-ray counts. This is the first result which has quantified the two gamma-ray components, and has opened a possibility to explore the physical processes operating in each part of the SNR. It is to be emphasized that the separation of the two gamma ray components was {achieved} for the first time only by including explicitly the ISM proton column density. The hybrid origin was theoretically modeled by \citetalias{2010ApJ...708..965Z} (see their Figure 14). The present results indicate that the hadronic component is more significant than shown by \citetalias{2010ApJ...708..965Z} which assumed a small target mass of the ISM. {In fact, \citetalias{2010ApJ...708..965Z} assumed the interacting molecular mass to be 100--1000\,$M_{\odot}$ which comprises only two individual molecular clouds C and D \citep{2008AIPC.1085..104F}. The total ISM mass associated with the SNR, however, {is} amounts to $\sim$20,000\,$M_{\odot}$ by summing up all molecular and atomic clouds interacting with the SNR \citepalias{2012ApJ...746...82F}.}

The principal aim of \citetalias{2012ApJ...746...82F} was to clarify if the target protons in the hadronic process in RX~J1713 show good correspondence with the interstellar protons including \ion{H}{1} in addition to H$_2$. The new feature was the inclusion of \ion{H}{1}, and the spatial correspondence is the necessary condition of the hadronic gamma rays. It was also shown by \citetalias{2012ApJ...744...71I} that the magnetic field is likely as strong as 100~mG or higher, making the leptonic component unfavorable. Their azimuthal angle distribution of the gamma rays and ISM seems to be matched without invoking an additional contribution\footnote{We note that in their Figure 8(b) large deviations are found at 15$\arcdeg$ and 165$\arcdeg$ in azimuth angle in their Section 3.5.2. 115$\arcdeg$ is to be corrected to 15$\arcdeg$, see the erratum.}. Accordingly, \citetalias{2012ApJ...746...82F} concluded that the {gamma rays are mainly due to the hadronic} origin, while the contribution of the leptonic component was not excluded. The subsequent works on CTA simulations of the hadronic and leptonic origins pursued this issue and argued for a possibility of resolving the degeneracy originating by the shell-like distribution of the ISM and nonthermal X-rays \citep{2017ApJ...840...74A}. In the other two TeV gamma ray SNRs, HESS~J1731$-$347 and RCW~86, the ISM and the nonthermal X-rays show significantly different spatial distributions, allowing one to separate the two origins \citep{2014ApJ...788...94F,2019ApJ...876...37S}, while the other TeV gamma ray SNR RX~ J0852.0$-$4622 shows shell-like distributions of the ISM and nonthermal X-rays, making  a case similar to RX~J1713. In order to shed more light on the issue we applied the present method to the previous data H.E.S.S.07 used by \citetalias{2012ApJ...746...82F}, and find that the ratio of the hadronic component and the leptonic components is $\sim$($72\pm11$) : ($28\pm10$) in gamma-ray count, which is similar to the present results. The $3\sigma$ significance of the leptonic component might place it at a marginal level, while the low spatial resolution and the small number of the data pixels would make the analysis of H.E.S.S.07 crude at best.


\subsection{Possible Suppression of the Gamma Rays by the 2nd Order Effects} \label{2ndF}
{A further} issue is the possible suppression of the gamma rays by the 2nd order effects that may make the gamma rays deviate from a simple proportionality to $N_{\rm p}$ or $N_{\rm x}$. Such effects include the shock cloud interaction and the penetration depth. Some other effects like secondary particle acceleration may be operating as suggested by \citet{2004vhec.book.....A}, while these effects are not yet fully investigated$/$confirmed theoretically or observationally; such mechanisms include {the 2nd order Fermi acceleration \citep{1949PhRv...75.1169F}}, magnetic reconnection in the turbulent medium \citep{2012PhRvL.108m5003H}, reverse shock acceleration \citep[e.g.,][]{2001SSRv...99..305E}, {and} non-liner effect of DSA \citep[e.g.,][]{2001RPPh...64..429M}. We shall not discuss them in the present paper. 

\citetalias{2012ApJ...744...71I} made MHD numerical simulations of the shock front interacting with the clumpy ISM. Recent MHD simulations of the dense cores overtaken by the shock by \citet{2019MNRAS.487.3199C} for more realistic core density higher than that of \citetalias{2012ApJ...744...71I} confirmed the results of \citetalias{2012ApJ...744...71I}. The interaction produces highly turbulent velocity field around the dense cores, which amplifies the magnetic field up to 100\,$\mu\textrm{G}$--1\,mG. The effects are observationally confirmed by the rim-brightened X-rays for {a} typical radius of the dense cores 0.5\,pc by \citet{2013ApJ...778...59S} through a detailed comparative study of the ISM and X-rays.
The $\sim$100\,$\mu\textrm{G}$ $B$ field will deplete CR electrons by synchrotron cooling and can lead to suppression of the leptonic gamma rays.

In the hadronic origin, the gamma rays are not always perfectly proportional to the ISM column density because CRs cannot penetrate freely into the dense cores as pointed out by \citet{2007Ap&SS.309..365G} and \citetalias{2012ApJ...744...71I}. The expression of the penetration depth of CR protons into dense cores is given as follows \citepalias{2012ApJ...744...71I};
\begin{eqnarray}
l_{\mathrm {pd}}\;({\mathrm {pc}}) & = & 0.1\;\eta^{1/2}\left( \frac{E}{10\,{\mathrm {TeV}}} \right)^{1/2} \nonumber \\
 & & \times \left( \frac{B}{100\,{\mu}G} \right)^{-1/2}\left( \frac{t_{\mathrm {age}}}{10^{3}\,\mathrm {yr}} \right),
\label{equ:7}
\end{eqnarray}  

where $l_{\rm {pd}}$ is the penetration depth of cosmic ray{s}, $\eta$ is the CR gyro-factor, $E$ is the cosmic-ray energy, $B$ is the magnetic field strength, and $t_{\rm {age}}$ is the age of the SNR. If we adopt $E = 10$\,TeV\footnote{{The gamma-ray images used for the represent paper correspond to the CR proton energy $\sim$3--20 TeV, which is roughly an order of magnitude higher than the lower energy threshold of the images.}}, $t_{\rm {age}} = 1600$\,yr, $B = 100\,\mu\textrm{G}$ \citep{2019MNRAS.487.3199C}, and $\eta = 1$ {\citep[e.g.,][]{2008ApJ...685..988T,2019ApJ...877...96T}}, we estimate $l_{\rm {pd}} = 0.13$\,pc. The dense cores in RX~J1713 have a typical radius of $\sim0.5$\,pc \citep{2010ApJ...724...59S, 2013ApJ...778...59S} and $l_{\rm {pd}}$ is small enough to affect the CR injection into the dense cores. {Here, t}he volume filling factor of the dense cores {in {RX~J1713} is} $\sim$10$\%$ {by assuming 22 molecular clouds with typical radii of $\sim$1.4 pc and the SNR shell radius of 9 pc \citep{2013ApJ...778...59S}. T}he gamma ray suppression is {therefore} not likely to be dominant overall.

Recently, \citet{2020ApJ...904L..24S} revealed high resolution distribution of CO with ALMA toward part of the peak D cloud, and small-scale (0.1--0.02\,pc) clumpy distribution has been revealed. If such a highly clumpy distribution is common in the ISM, CRs may more easily penetrate into dense ISM and may increase the hadronic gamma rays more than inferred from unresolved ISM distribution. So, it will be important to resolve the ISM at a sub-pc scale in order to better understand the hadronic gamma rays.

In either case, the suppression is supposed to appear at high $N_\mathrm{g}$, and the slight decrease of ${\Delta N_\mathrm{g}}$ {in only three pixels at around 2.5 $\sigma$ level} in Figure \ref{fig:3derr} might be caused by these effects. {The three pixels are indicated by green lines in Figure \ref{fig1}(a).} This issue is to be clarified by future more sensitive observations. Overall, the {present methodology} appears to describe the general behaviour {of $N_\mathrm{g}$}.

\subsection{Prospect to Improve Accuracy of the ISM Column Density} \label{dis:acc_ISM}
The present work used the ISM distribution consisting of both atomic and molecular hydrogen \citepalias{2012ApJ...746...82F}. A problem we faced is that the accuracy of estimating N(\mbox{\ion{H}{1}}) is limited by the \ion{H}{1} self-absorption. By new identification of possible self-absorption \ion{H}{1}, we excluded {18} pixels from the analysis, where self-absorption effects are suspected. Since the accuracy in correction for the self-absorption is limited by the uncertainty in the assumed \ion{H}{1} background emission, it is impossible to achieve accuracy better than $\sim$20\% for a correlation analysis if self-absorption is significant. If the correction is not adequate in some pixels of the present analysis, it leads to underestimate the ISM column density and $\hat{N}_{\rm g}$ as well. Then, $\Delta N_{\rm g}$ becomes overestimated in Figure \ref{fig:3derr}. {The four pixels with deviation more than 2.5$\sigma$ in a range of $\hat{N_\mathrm{g}}=1.0$--2.0 may show such a case.} In order to correct such inaccuracies, we need to develop a new tracer both for \ion{H}{1} and H$_2$. \citet{2012ApJ...759...35I} made numerical simulations of magneto-hydrodynamics of colliding \ion{H}{1} flows and traced the \ion{H}{1} evolution leading to H$_2$ cloud formation. By synthetic observations of the numerical results of \citet{2012ApJ...759...35I}, \citet{2018arXiv181102224T} and \citet{2018ApJ...860...33F} show that the sub-mm transition of neutral carbon \ion{C}{1} is a good tracer of the whole ISM protons (\ion{H}{1}$+$H$_2$) in a density regime around $10^3$ cm$^{-3}$ \citep[see Figure 8 of][]{2018arXiv181102224T}. We suggest that \ion{C}{1} serves as a useful probe of ISM protons in future and may become a promising tool in a ISM study of SNRs. This is particularly relevant in the CTA era, where more SNRs in the Galactic plane, subject to strong \ion{H}{1} absorption, will be discovered.

\subsection{The SNR Evolution in the Last 1600\,yr} \label{dis:SNRevo}
Once a SNe occurs, CRs are accelerated by the DSA in the SNR. Non-thermal X-rays are emitted by the synchrotron mechanism, and the gamma rays by the p-p reaction or the {inverse Compoton} scattering. A general picture of the hadronic and leptonic gamma rays from a non-thermal X-ray SNR is that the gamma rays are proportional to either $N_{\rm p}$ or $N_{\rm x}$. In RX~J1713, the CRs are perturbed by the interaction with the ISM. The distribution of the ISM is highly clumpy with the dense cores having a small volume filling factor of $\sim$10\% in RX~J1713. The distribution was determined prior to the SNe by the cloud self-gravity which may involve star formation as in the peak C \citep{2010ApJ...724...59S} and the environmental effects including the progenitor's stellar winds over Myr. The theoretical studies of the interaction show that the gamma rays and X-rays are significantly affected by the dense cores (\citealt{2007Ap&SS.309..365G}; \citetalias{2012ApJ...744...71I}; \citealt{2019MNRAS.487.3199C}). The secondary acceleration of CRs might also affect the X-rays as suggested by a spectral analysis of X-rays by \citet{2015ApJ...799..175S}, whereas there is no established picture on the acceleration.

The present hybrid picture is qualitatively consistent with the composite scenario by \citetalias{2010ApJ...708..965Z}. Naturally, the fraction of the leptonic gamma rays increases where the ISM density is low. In the other three TeV gamma ray SNRs, HESS~J1731$-$347 (HESS~J1731), RX~J0852.0$-$4622, and RCW86, spatial correspondence between the gamma rays and the ISM has been shown and the hadronic gamma rays are suggested to be dominant \citep{2014ApJ...788...94F, 2017ApJ...850...71F, 2019ApJ...876...37S}. In two of them, HESS~J1731 and RCW86, in a minor portion of the SNRs where the gamma-ray distribution shows correspondence with the nonthermal X-rays but not with the ISM, it is suggested that the gamma rays are mostly of the leptonic origin \citep{2014ApJ...788...94F, 2019ApJ...876...37S}. RX~J1713 is a more complicated case where the non-thermal X-rays are heavily overlapped with the gamma rays and are not easily separable at the current resolution. The methodology proposed in the present work has opened a new tool to {quantify} the leptonic and hadronic gamma rays and will be applicable to the other SNRs. The present paper lends support for the vital role of the ISM in the production of the HE and VHE energy radiations in a SNR, confirming that the inclusion of the ISM in an analysis is crucial for our understanding the origin of the gamma rays.

{Furthermore, numerical simulations of the gamma ray SEDs in RX~J1713 have been made by \citetalias{2010ApJ...708..965Z} as well as the other authors (see references in \citetalias{2010ApJ...708..965Z}). These simulations assumed the ratio of relativistic electrons to protons $K_\mathrm{ep}$ of 10$^{-2}$ to 10$^{-3}$ and obtained results which are not inconsistent with the present result. This suggests that the assumed values above satisfy the present requirement of the gamma-ray and X-ray results in addition to the ISM data, whereas it seems that better statistics of the gamma-ray data, which will be provided in the CTA era, would enable us to better constrain $K_\mathrm{ep}$.}
 
\section{Conclusions} \label{sec:con}

In order to pursue the origin of the gamma rays in the SNR RX~J1713.7$-$3946 (RX~J1713) we have carried out a comparative study of the H.E.S.S. TeV gamma rays with the ISM protons and the non-thermal X-rays. The main conclusions of the present work are summarized as follows.

\begin{enumerate}

\item{{We propose a new methodology which assumes that the gamma-ray counts $N_{\rm g}$ is expressed by a linear combination of two terms; one is proportional to the ISM column density Np and the other proportional to the X-ray count $N_{\rm x}$. By fitting the expression to the data pixels, we find that the gamma-ray counts are well represented by a tilted flat plane in a 3D space of $N_{\rm p}$-$N_{\rm x}$-$N_{\rm g}$. This plane illustrates that the total gamma rays $N_{\rm g}$ increases with $N_{\rm p}$ and $N_{\rm x}$, respectively, which is consistent with the hybrid picture. The fitting is robust with the goodness of fitting $R^2=0.92$ close to a complete fit with $R^2=1.0$. The results show that the hadronic and leptonic components occupy (58--70)\% and (25--37)\% of the total gamma rays, respectively. The two components have been quantified for the first time by the methodology. The hadronic component is typically by a factor of $\sim$2 greater than the leptonic component, especially as yielded by the highest resolution dataset at $>$2 TeV. The dominance of the hadronic gamma rays reflects the massive ISM of $\sim10^4$~$M_{\odot}$, target protons in the p-p reaction, associated with the SNR (\citetalias{2012ApJ...746...82F}), lending further support for the acceleration of the cosmic-ray protons.}}
\item{The theoretical broad band fit to the gamma-ray and X-ray spectra shows it likely that the leptonic gamma rays are existent in RX~J1713 as well as the hadronic gamma rays. The new methodology has opened a possibility of a pixel-resolved analysis by the broad-band gamma-ray and X-ray spectral fitting as a next step. \citetalias{2012ApJ...746...82F} showed that the ISM distribution has a good spatial correspondence to the gamma rays, and suggested that it supports the production of the hadronic gamma rays. The contribution of the leptonic gamma rays was, however, not quantified in the previous observational studies in part due to the degeneracy of the two components of the similar shell-like distribution (\citealt{2007AA...464..235A}; \citealt{2018A+A...612A...6H}; \citetalias{2012ApJ...746...82F}). The nonsignificant contribution of the leptonic component in \citetalias{2012ApJ...746...82F} may be ascribed to the low photon counts of the previous H.E.S.S. dataset in 2007 and the low resolution $\sim$10$\arcmin$ employed by \citetalias{2012ApJ...746...82F}.}

\item{{There is a marginal hint that the gamma rays are suppressed at high gamma-ray counts which may be ascribed to the second order effects including the shock-cloud interaction and the penetration-depth effect. The shock-cloud interaction excites turbulence toward the dense cores and amplifies the turbulent magnetic field up to 100 $\mu$G. The strong field deceases the CR electrons by the synchrotron cooling, leading to suppress the leptonic gamma rays by the inverse Compton scattering. In addition, the CR protons cannot penetrate into the dense cores because of the limited penetration depth around the dense cores where the turbulent magnetic field reduces the CR diffusion. This reduces the hadronic gamma rays toward the dense cores. These two effects will suppress the gamma rays toward the dense cores. In the CTA observations at high resolution, the two effects may become more significant than shown here.}}

\item{{The present analysis employed the \ion{H}{1} data corrected for the self-absorption, whereas in some region of the SNR the accuracy in the correction may be limited due to difficulty in estimating the background \ion{H}{1} emission. This can be an obstacle in future applications of the methodology to distant SNRs in the Galaxy, where stronger \ion{H}{1} absorption is expected. We suggested that the \ion{C}{1} sub-mm transition can be better used as a promising tracer of the whole ISM protons of both atomic and molecular hydrogen, which may help to improve the accuracy in the ISM column density. The development of such new probes will be crucial in future studies of the gamma ray origins. ALMA is obviously a powerful high resolution instrument of \ion{C}{1} observations and NANTEN2 will provide a large-scale view of \ion{C}{1} distribution by determining a X$_{\mathrm{CI}}$ factor for converting $W$(\mbox{\ion{C}{1}}) into $N_{\mathrm{H}}$. The combination of the two with CTA will greatly facilitate the accurate measurements of the interstellar protons.}}

\end{enumerate}

\acknowledgments
{We are grateful to the anonymous referee for his/her valuable comments on an earlier version of the manuscript that greatly improved the manuscript.} The NANTEN project is based on a mutual agreement between Nagoya University and the Carnegie Institution of Washington (CIW). We greatly appreciate the hospitality of all the staff members of the Las Campanas Observatory of CIW. We are thankful to many Japanese public donors and companies who contributed to the realization of the project. This work is financially supported by a grant-in-aid for Scientific Research (KAKENHI, Nos. 24224005, 15H05694, 19H05075, 20H01944, and 21H01136) from MEXT (the Ministry of Education, Culture, Sports, Science and Technology of Japan).
\facilities{ALMA, H.E.S.S., NANTEN}\\

\appendix

\section{{Distance determination of RX~J1713}}\label{sec:app:distance}

{In the Galactic plane, accurate determination of distance to a SNR generally involves large uncertainty, since SNRs have no clear correlation between the observed size and radio brightness. It is often the case that the kinematic distance offers a reliable distance once an association between a SNR and the interstellar clouds is established. In case of RX~J1713, there seems to be some issues concerning the method of distance estimate in the literature. We will review the previous work, and clarify the background of the present study.}

{The SNR RX J1713 was first identified by \cite{1996rftu.proc..267P} by ROSAT at $1'$--$6'$ resolution in the X-ray energy range of 0.1--2.5~keV, and the X-ray spectrum was interpreted to be thermal. They derived X-ray absorption column density to be (3--$12) \times 10^{21}$ cm$^{-2}$ for an effective resolution of $\sim17'$. \cite{1997PASJ...49L...7K} made X-ray observations with ASCA in a range of 0.7--10.0~keV, which covered the northwestern shell with the FoVs of $22' \times 22'$ square (SIS) and a circular shape with a diameter of $50'$ (GIS). These authors derived the average absorption column density of the shell region to be (4.8--10.1)$ \times 10^{21}$ cm$^{-2}$ from X-rays. The absorption occurs at low energy less than 1~keV, and the two works yielded similar average column densities. Pferffermann and Aschenbach estimated the distance to be 1~kpc by fitting a Sedov model. \cite{1997PASJ...49L...7K} also derived a distance of 1~kpc of the SNR corresponding to the average column density. These works assumed the gas distribution is uniform in the Galactic disk, and referred to the total column density to the Galactic center, $6 \times 10^{22}$ cm$^{-2}$, which is roughly ten times higher than that derived by them. This assumption however turned out to be incorrect in two ways later. First, the gas in the disk toward RX~J1713 is far from uniform and there is a significant decrease of hydrogen column density, which is seen as a hole of CO (see below). Second, most of the absorption is caused within the SNR and is not proportional to distance; \cite{2015ApJ...799..175S} showed that the X-ray absorption is divided into that within the SNR and that in a single foreground discrete cloud, where their ratio is $7:3$ \citep[see Figure 7a of][]{2015ApJ...799..175S}.}

{Subsequently, \cite{1999ApJ...525..357S} found that the interstellar gas has a hole toward the SNR by chance, raising a doubt on an assumption of uniform column density. These authors inspected CO distribution in the CfA survey at $9'$ resolution \citep{1989ApJS...71..481B} and found that clouds at $-94$--$-69$~km~s$^{-1}$ show spatial correspondence with the SNR. For these clouds, they derived a kinematic distance of 6~kpc by using the Galactic rotation curve. However, \citetalias{2003PASJ...55L..61F} used the NANTEN CO survey and identified that another cloud at $-10$~km~s$^{-1}$ shows good correspondence with the SNR at a higher resolution of 2\farcm6 with a higher sensitivity. These authors thus derived a kinematic distance to be 1~kpc and \cite{2005ApJ...631..947M} supported the distance by presenting a full account of the NANTEN CO data in the region. Soon after, \cite{2004A&A...427..199C} confirmed the correspondence of the spatially varying X-ray absorption with the \ion{H}{1} and CO, and argued for a distance of 1.3~kpc. The readers are able to refer to a full account of the X-ray absorption, which varies mainly over a large range of 0.4--$1.4 \times 10^{22}$~cm$^{-2}$ in column density, in \cite{2015ApJ...799..175S} based on the Suzaku X-ray data. The distance is supported by a number of subsequent works on the association of the $-10$~km~s$^{-1}$ cloud with the SNR \citep[e.g.,][]{2009A&A...505..157A,2010ApJ...724...59S,2013ApJ...778...59S,2015ApJ...799..175S,2020ApJ...900L...5T}, and seems to be confirmed by the recent GAIA data \citep{2021NatAs.tmp...79L}.}

{In summary, the distance of RX~J1713, 1~kpc, is established based on the association of the $-10$~km~s$^{-1}$ cloud with the SNR (\citetalias{2003PASJ...55L..61F}; \citealt{2004A&A...427..199C,2005ApJ...631..947M}). A similar estimate of 1~kpc by the early X-ray works \citep[e.g.,][]{1997PASJ...49L...7K} was fortuitous due to an incorrect assumption of uniform column density in the Galactic disk.}

\begin{figure*}[]
\begin{center}
\includegraphics[width=\linewidth,clip]{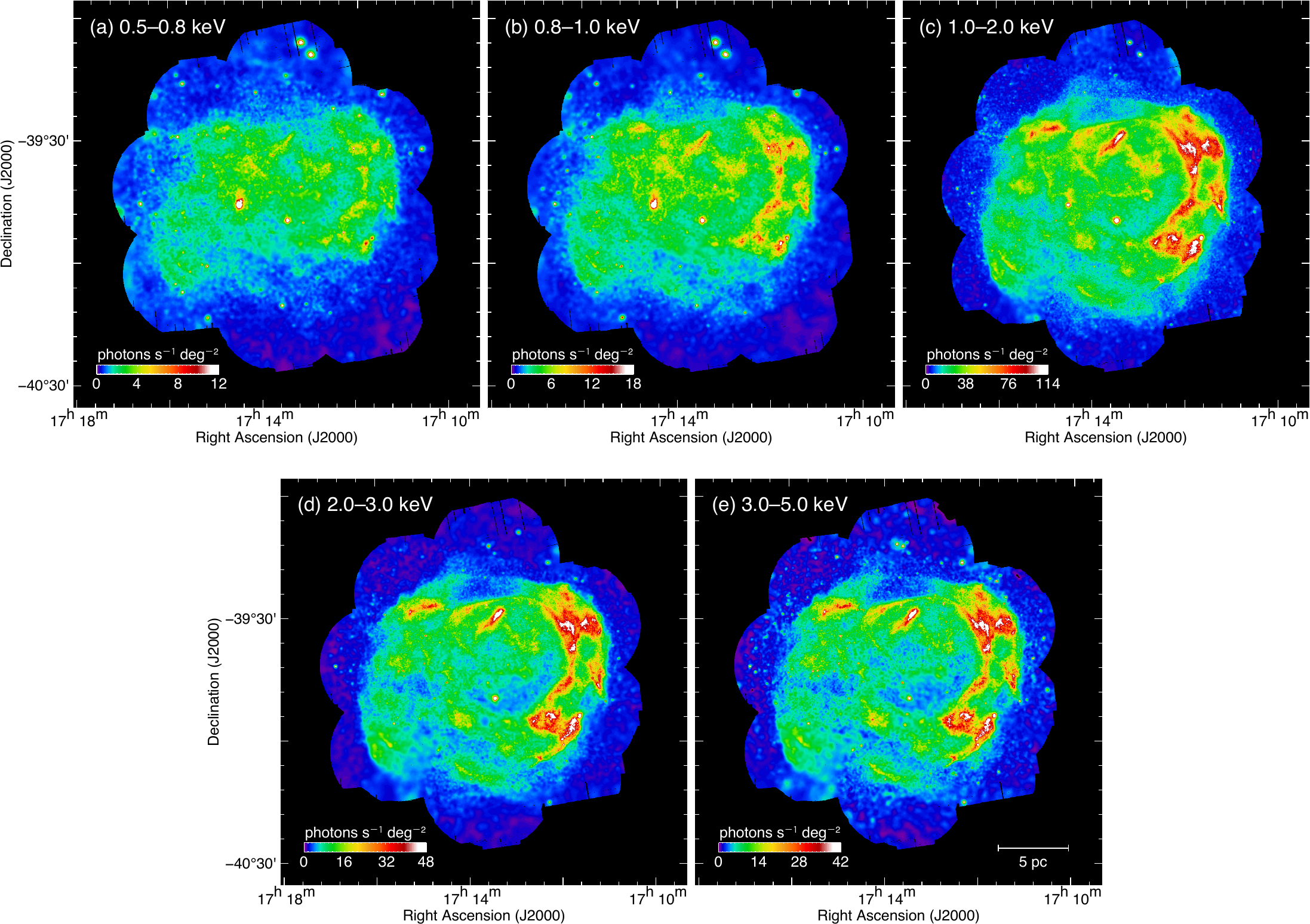}
\caption{Exposure-corrected background-subtracted X-ray flux images in energy bands of (a) 0.5--0.8\,keV, (b) 0.8--1.0\,keV, (c) 1.0--2.0\,keV, (d) 2.0--3.0\,keV, and (e) 3.0--5.0\,keV. The scale bar is also shown in the bottom right corner in Panel (e).}
\label{fig:app1}
\end{center}
\end{figure*}%
\begin{figure}[]
\begin{center}
\includegraphics[width=12 cm,clip]{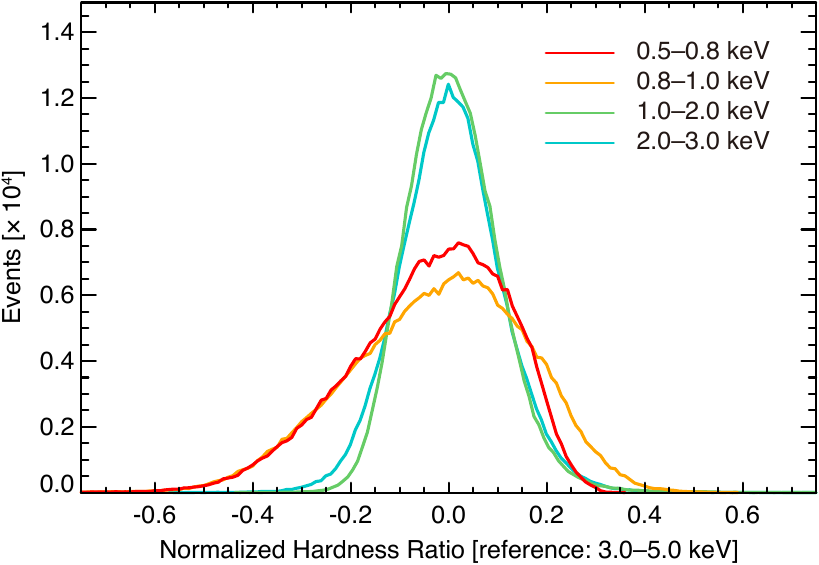}
\caption{Histograms of {normalized} hardness ratios respect to the energy band of 3.0--5.0~keV. The data was extracted from inside the shell eclipse centered at ($\alpha_\mathrm{J2000}$, $\delta_\mathrm{J2000}$) $=$ ($17^\mathrm{h}13^\mathrm{m}42\fs0$, $-39\arcdeg45\arcmin36\farcs0$), whose semimajor and semiminor radii are 0\fdg55 and 0\fdg48 with a position angle of $-51\arcdeg$.}
\label{fig:app2}
\end{center}
\end{figure}%

\section{X-ray absorption by the ISM}\label{sec:app}

It is well known that X-ray emission can be absorbed by the ISM which is located between the X-ray emitting source and observer. Considering the energy-dependent photoionization cross section of the ISM, the X-ray absorption becomes significant in the soft X-ray band generally below 1\,keV when the ISM column density is $\sim$10$^{22}$\,cm$^{-2}$ \citep[e.g.,][]{1992ApJ...400..699B}. Since the typical absorbing column densities of {RX~J1713} is $\sim$0.4--$1.0 \times 10^{22}$\,cm$^{-2}$ by X-ray spectroscopies, soft-band X-ray images might be affected by the interstellar absorption \citep[e.g.,][]{2004A&A...427..199C,2015ApJ...799..175S,2018PASJ...70...77O}. In this section, we shall test which energy bands of X-ray images are free for the interstellar absorption in {RX~J1713}.

Figure \ref{fig:app1} shows exposure-corrected background-subtracted X-ray flux images in energy bands of 0.5--0.8, 0.8--1.0, 1.0--2.0, 2.0--3.0, and 3.0--5.0\,keV. We find that three X-ray images of 1.0--2.0, 2.0--3.0, and 3.0--5.0\,keV show similar spatial distributions to each other. On the other hand, 0.5--0.8 and 0.8--1.0\,keV images are significantly different with the others especially for the southern part of the shell. The X-ray depression in soft-band images is likely caused by the interstellar absorption of foreground local clouds, which was previously mentioned by \citet{2005ApJ...631..947M} and \citet{2015ApJ...799..175S}.

To evaluate the absorption effect in the soft band X-ray images, we calculated hardness ratios HR which provides a photometric color index by a following equation using high-energy band image $H$ and low-energy band image $S$;
\begin{equation}
\mbox{HR} = (H - S) / (H + S).      
\label{equ:a1}
\end{equation}
In the present study, we fixed high-energy band image $H$ to 3.0--5.0\,keV image for the reference {and normalized the hardness ratio whose Gaussian center will be adjusted to zero}.

Figure \ref{fig:app2} shows histograms of {normalized} hardness ratios within the SNR shell for each low-energy band images. We find the dispersions of histograms for 1.0--2.0 and 2.0--3.0\,keV are narrow, whereas the other two soft bands below 1~keV show wider dispersions. The dispersion (full-width half-maximum) obtained by a Gaussian fitting is 0.37 for 0.5--0.8\,keV, 0.45 for 0.8--1.0\,keV, 0.22 for 1.0--2.0\,keV, and 0.22 for 2.0--3.0\,keV. Since the dispersion reflects intensity variations of an X-ray image due to the interstellar absorption, we conclude that X-ray images of 1\,keV or higher energy could be considered as the absorption free or neglectable in {RX~J1713}. We therefore used the 1.0--5.0\,keV image for the correlation study.

\section{multiple linear regression}\label{sec:multiple_regression}
We have made a weighted least-squares estimation of the coefficients in Eqation (\ref{equ:4}) and their standard error, goodness-of-fit tests, and {red}{a diagnostic of multicollinearity}, according to Chapters 6, 10 and 11 of \citet{Kutner2005}.

\subsection{Estimation of the Regression Coefficients}\label{subsec:multiple_regression_coeff}

Equation (\ref{equ:4}) can be rewritten in a matrix form
\begin{equation}
\left(
	\begin{array}{c}
		N_{\mathrm{g}, 1} \\	
		\vdots \\
		N_{\mathrm{g}, n}	
	\end{array}
\right)=
\left(
	\begin{array}{cc}
		N_{\mathrm{p}, 1} & N_{\mathrm{x}, 1} \\	
		\vdots & \vdots \\
		N_{\mathrm{p}, n} & N_{\mathrm{x}, n} 	
	\end{array}
\right)
\left(
	\begin{array}{c}
		a \\	
		b
	\end{array}
\right)+
\left(
	\begin{array}{c}
		\varepsilon_{1} \\
		\vdots \\
		\varepsilon_{n}
	\end{array}
\right), 
\end{equation}
or more simply
\begin{equation}
\mathbf{y}=\mathbf{Xb}+\boldsymbol{\varepsilon}, 
\end{equation} 
where $\mathbf{y}=(\begin{array}{ccc} N_{\mathrm{g}, 1} & \cdots & N_{\mathrm{g}, n} \end{array})^\mathrm{T}$ is a vector of $n$ observations on the dependent variable $N_\mathrm{g}$ and the superscript T means transpose, 
\begin{displaymath}
\mathbf{X}=\left(
	\begin{array}{cc}
		N_{\mathrm{p}, 1} & N_{\mathrm{x}, 1} \\	
		\vdots & \vdots \\
		N_{\mathrm{p}, n} & N_{\mathrm{x}, n} 
	\end{array}
\right)
\end{displaymath}
is an $n\times 2$ matrix where we have observations on two explanation variables, $N_\mathrm{p}$ and $N_\mathrm{x}$, for $n$ observations, $\mathbf{b}=(\begin{array}{cc} a & b \end{array})^\mathrm{T}$ is a vector of unknown coefficients which we want to estimate and $\boldsymbol{\varepsilon}=(\begin{array}{ccc} \varepsilon_{1} & \cdots & \varepsilon_{n} \end{array})^\mathrm{T}$ is a vector of disturbances.

An estimate of the coefficients $\hat{\mathbf{b}}$ which minimize the sum of squared residuals $\boldsymbol{\varepsilon}^\mathrm{T}\boldsymbol{\varepsilon}$ is given as
\begin{equation}\label{eqn:vector_estimated_b}
\hat{\mathbf{b}}=(\mathbf{X}^\mathrm{T}\mathbf{WX})^{-1}\mathbf{X}^\mathrm{T}\mathbf{Wy},
\end{equation}
where an $n \times n$ diagonal matrix
\begin{displaymath}
\mathbf{W}=\left(
	\begin{array}{ccc}
		w_{1} & & \\
		 & \ddots & \\
		 & & w_{n}
	\end{array}
\right)
\end{displaymath}
is the weight matrix, $w_{i}=\sigma(N_{\mathrm{g}, i})^{-2}$ ($i=1\cdots n$).
Here, we do not take into account $\sigma(N_\mathrm{p})$ and $\sigma(N_\mathrm{x})$ because they are relatively small; $\sigma(N_\mathrm{p})/N_\mathrm{p}= 10^{-2}$ and $\sigma(N_\mathrm{x})/N_\mathrm{x}= 10^{-2}$--$10^{-3}$, while $\sigma(N_\mathrm{g})/N_\mathrm{g}=10^{-1}$.
The standard error of the estimated coefficients, $\sigma_{a}$ and $\sigma_{b}$, are given as the square root of the diagonal elements of a matrix
\begin{equation}\label{eqn:standard_error_matrix}
\frac{(\mathbf{y}-\mathbf{\hat{y}})^\mathrm{T}\mathbf{W}(\mathbf{y}-\mathbf{\hat{y}})} {n-m} (\mathbf{X}^\mathrm{T}\mathbf{WX})^{-1},
\end{equation}
where $\mathbf{\hat{y}}=\mathbf{X\hat{b}}=(\begin{array}{ccc} \hat{N}_{\mathrm{g}, 1} & \cdots & \hat{N}_{\mathrm{g}, n}\end{array})^\mathrm{T}$ is a vector of predicted values of $N_\mathrm{g}$ and $m=2$ is the number of explanation variables ($N_\mathrm{p}$ and $N_\mathrm{x}$).

\subsection{Goodness-of-Fit Measure and $F$-Test}\label{subsec:goodnes_of_fit}
Taking into account the particularities of regression through the origin (briefly summarized in Appendix A of \citealt{legendre2009independent}), we calculate the coefficient of determination\footnote{According to \citet{kvaalseth1985cautionary}, there are eight types of definitions of $R^{2}$. Equation (\ref{eqn:R2}) corresponds to Equation (8) of this author.} as
\begin{equation}\label{eqn:R2}
R^{2}=\displaystyle{\sum_{i=1}^{n} (\hat N_{\mathrm{g}, i})^{2}}\bigg/\displaystyle{\sum_{i=1}^{n}(N_{\mathrm{g}, i})^{2}}
\end{equation}
and $F$-statistic associated with $R^{2}$ as
\begin{equation}\label{eqn:f-statistic}
F=\frac{R^{2}/m}{(1-R^{2})/(n-m)}.
\end{equation}
$R^{2}$ indicates how the variance in the dependent variable ($N_\mathrm{g}$) can be explained by those in the explanation variables.
It normally ranges from 0 to 1 and a high $R^{2}$ indicates a good fit.
The derived $F=337.0$ is much greater than the critical value of the $F$-distribution with ($m$, $n-m$) degrees of freedom, $F_{0.01}(2, n-2)=4.9$, therefore we can reject a null-hypothesis $a=b=0$ which means $N_\mathrm{g}$ is dependent on neither $N_\mathrm{p}$ nor $N_\mathrm{x}$, at a significance level of 1\%.

\subsection{Diagnostics of Multicollinearity}
{red}{
Multicollinearity is a well known problem which increases the variance of the coefficient estimates ($\sigma_{a}$ and $\sigma_{b}$ in the present study) and makes the estimates unstable.
It occurs when the explanation variables in the regression model are highly correlated.
A formal and widely accepted method for detectidng the presence of multicollinearity is use of variance inflation factor (VIF).
A VIF value greater than 10 is commonly used as an indicator of multicollinearity but a more conservative threshold of 5 or 3 is sometimes chosen.
}

{red}{
There are only two explanation variables ($N_\mathrm{p}$ and $N_\mathrm{x}$) in the regression model in the present study, the VIF is given by
\begin{equation}
\mathrm{VIF}=\frac{1}{1-r^{2}},
\end{equation}
where $r$  is the correlation coefficient between $N_\mathrm{p}$ and $N_\mathrm{x}$.
The correlation coefficient of $r=0.71$ -- 0.80 give VIF of 2.0 -- 2.8 which are in an acceptable range.
}

\bibliography{references}{}
\bibliographystyle{aasjournal}

\end{document}